\documentclass[lettersize,journal,twoside]{IEEEtran}
\usepackage{graphicx}
\usepackage{cite}
\usepackage{amsmath}
\usepackage{amsfonts}
\usepackage{mathtools}
\usepackage{booktabs}
\usepackage[table,xcdraw]{xcolor}
\usepackage{float}
\usepackage{xurl}
\usepackage{cuted}
\usepackage{etoolbox}
\AfterEndEnvironment{strip}{\leavevmode}
\usepackage{nicefrac}

\begin{document}

\title{Estimating Time Delays between Signals under Mixed Noise Influence with Novel Cross- and Bispectral Methods}
\author{Tin Jurhar$^{1,2}$, Franziska Pellegrini$^{3,4}$, Ana I. Nuñes del Toro$^{4,5}$, Tilman Stephani$^{6}$, Guido Nolte$^{7}$,\\Stefan Haufe$^{2,3,4,5,*}$
\thanks{This work was supported by the European Research Council (ERC) under the European Union’s Horizon 2020 research and innovation programme (Grant agreement No. 758985). }
\thanks{Work was conducted while the authors were with the following institutions. $^{1}$Radboud Universiteit, Nijmegen, The Netherlands. $^{2}$Physikalisch-Technische Bundesanstalt Berlin, Berlin, Germany. $^{3}$Charit\'e Universit\"atsmedizin Berlin, Berlin, Germany. $^{4}$Bernstein Center for Computational Neuroscience, Berlin, Germany. $^{5}$Technische Universität Berlin, 10623 Berlin, Germany. $^{6}$Max Planck Institute for Human Cognitive and Brain Sciences, Leipzig, Germany. $^{7}$University Medical Centre Hamburg-Eppendorf, Hamburg, Germany.}
\thanks{$^{*}$Corresponding author: {\tt\small haufe@tu-berlin.de}}%
}

\markboth{Preprint}{Jurhar \MakeLowercase{\textit{et al.}}: Estimating Time Delays between Signals under Mixed Noise Influence with Novel Cross- and Bispectral Methods}

\maketitle

\begin{abstract} 
A common problem to signal processing are biases introduced by correlated noise. When quantifying time delays between two signals, mixed noise introduces a bias towards zero delay in conventional delay estimates based on the cross- or bispectrum. Here we propose two novel time delay estimators that address these shortcomings: (1) A cross-spectrum based approach that relies on estimating the periodicity of the phase spectrum rather than its slope, and (2) a bispectrum based approach, bispectral antisymmetrization, which removes contributions from not just Gaussian but all independent sources. In a simulation study, we compare conventional and novel TDE approaches and resolve differences in performance with respect to noise Gaussianity and auto-correlation structure. As a proof-of-concept, we also perform TDE analysis on a neural stimulation dataset ($n=3$).
We find that antisymmetrization consistently outperforms conventional bispectral methods at low signal-to-noise ratios (SNR) and prevents spurious zero-delay estimates in all mixed-noise environments. Time delay estimation based on phase periodicity also improves signal sensitivity compared to conventional cross-spectral methods. These observations are stable with respect to the magnitude of the delay and the statistical properties of the noise.
\end{abstract}

\IEEEPARstart{T}{he} problem of time-delay estimation (TDE) has already been considered half a century ago \cite{Knapp1976}. Still, current TDE approaches are limited in their application to complex noise environments. Three main approaches to TDE can be identified: cross-correlation, cross-spectrum and bispectrum based analysis. They all have in common that they are generally not robust against artifacts introduced by mixed noise. Although bispectrum based TDE has previously been proposed to specifically suppress contributions from correlated Gaussian sources \cite{Nikias1988}, none of these methods perform optimally in non-Gaussian noise environments. 

Here we introduce two novel TDE approaches that are insensitive to noise mixing. We derive a cross-spectrum based approach that relies on estimating the periodicity of the phase spectrum rather than its slope, and a bispectrum based approach that incorporates bispectral antisymmetrization \cite{Chella2014} in order to suppress contributions from not just Gaussian but all independent sources. Through theoretical arguments and by means of simulations, we demonstrate that our methods are immune to linear and instantaneous mixing effects. 

Since efforts of cross- and bispectrum TDE developed independently, comparisons between the two approaches are still lacking. A more general aim of this study is, therefore, to provide a versatile testing framework within which one can inspect the behaviour of each method under controlled signal and noise conditions. Using this framework, we simulate time-delayed time series. We assess conventional phase-slope and bispectrum based TDE approaches alongside the proposed phase-periodicity and bispectral antisymmetrization approaches in unmixed and mixed noise environments. Additionally, we characterize potential differences in approaches behaviour with respect to noise Gaussianity and auto-correlation. We further propose to combine standard TDE approaches with a bootstrapping procedure, which allows one to discern between reliable and unreliable TDE outputs. As a proof of concept, we demonstrate the viability of our TDE approaches on real-life electrophysiological data where time-delays between sensor measurements are expected. 

We proceed in the following way: Section~\ref{sec:theory} (Theory) formally introduces the TDE framework and outlines the theory of cross-correlational, cross-spectral, and bispectral TDE approaches. Based on existing theory, the section further discusses settings in which these methods are expected to either perform well or fail, with a particular focus on noise mixing. Then, we describe our novel cross-spectral phase-periodicity based and antisymmetrized bispectrum based TDE approaches, and provide theoretical arguments for their insensitivity to mixed noise artifacts. 
Sections~\ref{sec:expGauss}~\&~\ref{sec:expcolor} build on our theoretical arguments and comprise two simulation studies, in which we validate of our novel TDE approaches. In Section~\ref{sec:expreal}, we test the viability of our TDE approaches on a real-life electrophysiological dataset ($n=3$) where prior knowledge about the delays between sensor measurements is available. The remaining sections are dedicated to discussing the relevance of our findings.

\section{Theory} \label{sec:theory}
\subsection{Formalization of the TDE Problem}
\noindent Given two discrete time series $X(t),Y(t)$, we consider them time-delayed with delay $\tau$ if they follow the relationship

\begin{align}
X(t) &= x(t) +n_X(t) \label{eqn:lagX}\\
Y(t) &= \beta x(t-\tau) +n_Y(t)\label{eqn:lagY},
\end{align}
\noindent where $\beta = \{-1,1\}$ determines the polarity of $x(t-\tau)$ in $Y(t)$. Here we refer to $x(t)$ as the signal, while $n_{X,Y}(t)$ are additive noise terms. Furthermore, we require $x(t),n_{X,Y}(t)$ to have zero mean. If we do not need to distinguish between signal and noise time series, we simply call them sources. Time-delay estimation then entails all approaches which aim to quantify the time lag $\tau$ \cite{Nikias1988}.

We further make the distinction between the noiseless, the unmixed noise, and the mixed  noise environment. We can formalize these distinctions by introducing parameters $\alpha$ and $\theta_{1,2}$, and by unifying Eqs.~\eqref{eqn:lagX} \& \eqref{eqn:lagY} into

\begin{align} 
\begin{bmatrix}
X(t) \\
Y(t) \end{bmatrix} = \alpha \begin{bmatrix}
x(t) \\ \beta x(t-\tau) \end{bmatrix} + (1-\alpha) \begin{bmatrix}
1 & \theta_1 \\
\theta_2 & 1 \end{bmatrix} \begin{bmatrix}
n_X(t) \\
n_Y(t) \end{bmatrix}. \label{eq:fullTDE} 
\end{align} 
\\
\noindent Here, $\alpha \in [0, 1]$ quantifies the relative contributions of signal and noise to the overall time series, and $\theta_{1,2}$ quantify the degree of noise mixing. We will study the noiseless case $\alpha = 1$ to understand the behaviour of each TDE approach under ideal conditions and the alternative case ($0 \leq \alpha < 1$) to address differences in TDE outcome between unmixed ($\theta_{1,2} = 0$) and mixed ($\theta_{1,2} \neq 0$) noise environments. Though our model allows for opposite polarity of channels $X(t),Y(t)$, we focus on the homopolar setting ($\beta=1$) in the present paper and discuss potential differences with respect to TDE outcome only where necessary. 

\subsection{Challenges of TDE under Mixed Noise Influence}

\noindent The problem of noise mixing in the context of delay estimation is well described in the signal processing literature \cite{Knapp1976,ZhaoZhen1984,Nikias1993} and has more recently resurfaced in fields such as neuroscience \cite{Stam2007}, pipe leakage detection \cite{Brennan2007}, and acoustics \cite{Faerman2022}. 

In neuroscience, making functional inferences from non-invasive electrophysiological data is especially difficult. Individual source currents are passively conducted through the head volume and ultimately combine to a net signal, which makes it difficult to draw inferences on a source level. Under the quasi-static approximation of Maxwell's equations, these neuronal source currents combine additively and instantaneously \cite{Salmelin2009}. Hence, current efforts in neurophysiology concentrate on developing robust methodologies that suppress signal contributions from linearly mixed sources\cite{Nolte2004,Nolte2005,Nolte2008,Vinck2011,Chella2014,Bastos2016,Winkler2016,Baillet2020,Pellegrini2023}. Along these lines, we here propose two novel approaches to overcoming biases caused by mixing effects in TDE.

We note that in Eq.~\eqref{eq:fullTDE}, the correlation between $n_X(t),n_Y(t)$ is intentionally limited to an interaction at time lag zero, to model, for example, volume conduction effects in the head. In the mixed-noise case, information between $X(t)$ and $Y(t)$ is thus shared from signal source $x(t)$ at lag $\tau$, and from noise sources $n_X(t),n_Y(t)$ at lag zero. Mixed-noise TDE can then be understood as the introduction of an additional zero-lag interacting component, and its challenge is to disambiguate non-zero lag signal interactions from zero-lag noise interactions. Additionally, further nonzero interactions could be introduced between $X(t)$ and $Y(t)$ by choosing $x(t),n_{X,Y}(t)$ to have nonzero auto-correlation.

In order to understand the shortcomings of current TDE approaches, we proceed in the following way: We first characterize cross-correlational, cross-spectral, and bispectrum based TDE behaviour in the noiseless case. This serves to provide an intuition for the workings of each method. We then move on to the unmixed noise case, where we demonstrate how uncorrelated noise contributions are accounted for by each approach.  This is followed by an outline of TDE behaviour in the mixed noise setting. Here we show how superpositions of zero-lag interactions between noise components and non-zero lag interactions between signal components affect TDE outcome. In the last part of this section, we introduce two novel TDE approaches and provide theoretical arguments as to why they are insensitive to zero-lag interactions.

\subsection{TDE in the Time Domain}
\noindent A straightforward way to perform TDE is to compute the cross-correlogram of the two time series \cite{Knapp1976}. It is defined as 

\begin{align} 
r(\rho) = E\left[ X(t) Y(t+\rho) \right],\label{eq:corr}
\end{align}

\noindent where $E\left[.\right]$ denotes the expectation over time. In practice, $X(t), Y(t)$ can be segmented into $n = 1,...,N$ epochs and $r(\rho)$ can be computed as the average epoch-wise cross-correlogram. We further define the cross-correlational TDE estimate as

\begin{align}
\tau_{\text{corr}} = \arg \underset{\rho}{\max}\ |r(\rho)|,
\end{align}

\noindent where ($|.|$) denotes absolute value. In the noiseless case, $\tau_{\text{corr}}$ corresponds to the underlying delay $\tau$: substituting Eq.~\eqref{eq:fullTDE} into Eq.~\eqref{eq:corr} for $\alpha = 1$ and $n_{X,Y} = 0$ yields

\begin{align}
r(\rho) = \alpha^2 \beta E\left[x(t) x(t-\tau+\rho) \right],
\end{align}

\noindent which is maximal at $\tau = \rho$ for $\beta =1$ and minimal at $\tau = \rho$ for $\beta =-1$. The same result can be derived for the unmixed noise case, i.e. where $\theta_{1,2} = 0$ (see Supplement, Eq.~\eqref{supp:corrunmixed}). Contributions from uncorrelated noise thus perfectly cancel out when taking the expectation and can be mitigated in a finite sample regime by choosing a reasonably large observation window. This makes cross-correlational TDE suitable for unmixed noise environments.

In the mixed noise environment, cross-correlational TDE is unreliable. Here we expect to observe two interactions, namely the time-delayed interaction of signals and the instantaneous interaction of noise. In the Supplement (Eq.~\eqref{supp:corrmixed}), we show that the additional noise interaction is indeed reflected in the cross-correlogram by two additional noise terms (Eq.~\eqref{corr_cross_terms}):
\begin{align}
\nonumber r(\rho) & =  \alpha^2 \beta E\left[x(t)x(t-\tau+\rho)\right]\ldots \\ 
\label{corr_cross_terms} +&\theta_2 E\left[ n_X(t) n_X(t+\rho) \right] + \theta_1 E\left[ n_Y(t)n_Y(t+\rho) \right]
\end{align}

\noindent Both of these terms will peak at $\rho=0$. This means that the maximal value of $r(\rho)$, and thus $\tau_{\text{corr}}$, will depend on the relative size of each peak, and not necessarily reflect the underlying delay. In case the noise component exhibits auto-correlation, this may introduce further peaks to the cross-correlogram. It can therefore be said that spurious zero-delay reports through cross-correlational TDE are to be expected in low SNR environments. 

\subsection{TDE in the Frequency Domain}

\noindent Alternatives to cross-correlation based TDE are cross- and bispectrum based TDE approaches, which exploit the effect induced by time-delays in the frequency domain. For frequency domain TDE, we make use of the Discrete Fourier transform 

\begin{align}\label{eq:fouriertransform}
FFT(X(t)) = F_X(f) = \sum^{T-1}_{t=0} X(t)\ e^{-\frac{i2\pi}{T} ft} 
\end{align}

\noindent of $X$ and its inverse

\begin{align}\label{eq:invfouriertransform}
IFFT\left(F_X(f)\right) = \frac{1}{T} \sum_{f=0}^{T-1} F_X(f)\ e^{\frac{i2\pi}{T}ft} = X(t)
\end{align}

\noindent for $t = 1,\ldots,T$, and similarly for time series $Y$. $F$ is thus complex-valued of the form $re^{i\varphi}$, where $r$ denotes the magnitude and $\varphi$ the phase of $X$ or $Y$ at each frequency component $f$. We will make use of the notation $|re^{i\varphi}|:= r$ and $\angle(re^{i\varphi}):= \varphi$; the latter operation corresponds to the 2-argument arctangent (arctan2) function. Note that according to this definition the number of  Fourier coefficients corresponds to the number of time steps of the input: $T_F = T$. 

\begin{figure}[htbp]
\centering
\includegraphics[width=3.6in]{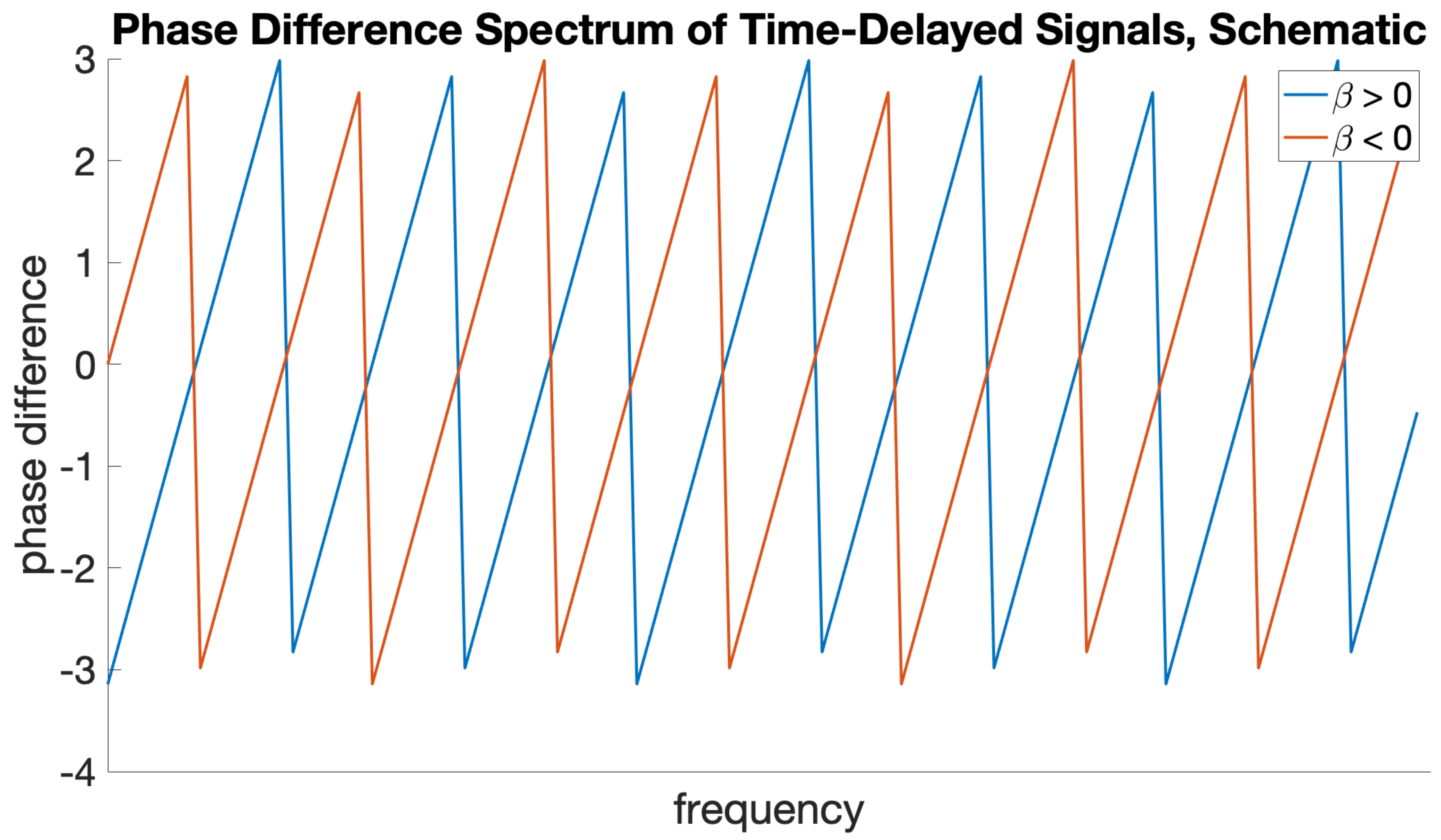}
\caption{Schematic representation of the phase difference spectrum of time-delayed signals for equal (blue) and opposite (orange) polarity. The two functions differ by a shift of $\pi$ along the frequency axis. }
\label{fig:phasediff}
\end{figure}

Fourier analysis is useful in the TDE framework considering that the notion of a time-delay can be simplified in the frequency domain: for signals with a single frequency component, a delay in the time domain is equivalent to a phase shift. For time-delayed sources $A$ and $B$ with multi-frequency components, the phase difference in each frequency component is proportional to the underlying frequency:

\begin{align}\label{eq:phasetodelay}
\varphi_A(f) - \varphi_B(f) = \!\!\! \mod(2\pi f \tau -\begin{Bsmallmatrix}\pi \text{, if $\beta = 1$} \\ 0 \text{, if $\beta = -1$} \end{Bsmallmatrix} ,2\pi)-\pi,
\end{align}

\noindent where the two alternative cases account for the phase shift elicited by channels with equal and flipped polarity. We further note that by nature of the two-argument arctangent function, the phase difference wraps around $[-\pi,\pi]$. The result is a sawtooth function whose slope and periodicity is fully characterized by the underlying delay $\tau$, and is schematically depicted in Figure~\ref{fig:phasediff}. Contrasting phase information of two time series is central to both cross- and bispectrum based TDE.

\subsection{Cross-Spectrum based TDE} \label{sec:cross spec TDE}
\noindent The cross-spectrum \cite{Brennan2007,Nolte2008,Faerman2022} is commonly used to estimate time delays as it inherently preserves phase relationships between two time series. Consider again segmented time series $X_s,Y_s$ and their Fourier transforms $F_X, F_Y$. If $X,Y$ are segmented, we can define the cross-spectrum as

\begin{align}
S_{XY}(f) &= \Big\langle F_X(f) F_Y^*(f)\Big\rangle = \Big\langle r_Xr_Y e^{i(\varphi_X-\varphi_Y)}\Big\rangle \label{eq:crossspectrum}
\end{align}

\noindent\cite{Nikias1993}, where $\langle.\rangle$ denotes the expectation across $N$ segments, and $(*)$ is the complex conjugate. The phase (difference) spectrum 

\begin{align}
P_{XY}(f) = \angle(S_{XY}(f)) \label{eq:phasespec}
\end{align}

\noindent isolates phase information of the cross-spectrum, which can then be used for cross-spectrum based TDE \cite{Piersol1981}. Note that $S_{XY}(f)$ and $P_{XY}(f)$ both have length $M_F=T$.

In cross-spectral TDE, prior information about the shape of the phase spectrum is exploited to arrive at a delay estimate. We first consider the noiseless case of two delayed time series: here, the phase spectrum, characterized by Eq.~\eqref{eq:phasetodelay}, is a linear function whose slope $b$ is proportional to the underlying delay $\tau$ \cite{ZhaoZhen1984}. We demonstrate this on a portion of the phase spectrum where phase wrapping is not expected to occur, i.e. the portion of Eq.~\eqref{eq:phasetodelay} within the modulo function:

\begin{align}
\nonumber    b(f) &= \frac{\Delta P_{XY}(f)}{\Delta f} = \frac{P_{XY}(f+1) - P_{XY}(f)}{(f+1)-f} \\
\nonumber    &= 2 \pi (f+1)\tau - \pi -  (2 \pi f \tau -\pi) \\
    &=  2 \pi f\tau +  2 \pi \tau -  2 \pi f\tau -\pi + \pi = 2\pi\tau,
\end{align}
\noindent for $\beta = 1$, and similarly for $\beta = -1$.

$P_{XY}(f)$ has several notable properties: with increasing $\tau$, $P_{XY}(f)$ exhibits (i) a steeper slope, resulting in (ii) faster excess of values $\pi$ or $-\pi$, which in turn causes (iii) more frequent phase wrapping and increased periodicity across the full spectrum. Furthermore, we acknowledge the phase-slope corresponding to a zero delay interaction will be $b_0 = 2\pi0=0$, and the resulting phase spectrum will thus be a flat line. 

One way to arrive at a  phase-slope estimate involves a least-squares linear fit of the estimator function $\hat P(f) = \hat bf$ over the unwrapped phase spectrum, which is then used to arrive at a delay estimate $\tau_{\text{phase-slope}}$:

\begin{align}
\nonumber b_\text{est} &= \arg\underset{\hat b}{\min} \sum_f \left(P_{XY}(f)-\hat P(f)\right)^2 \\
&= \arg\underset{\hat b}{\min} \sum_f \left(P_{XY}(f)-\hat bf\right)^2, \\
\tau_\text{phase-slope} &= \frac{b_\text{est}}{2\pi},
\end{align}

\noindent where $\hat b$ is sampled across a sufficiently large range of values. Interestingly, \cite{Brennan2007} show that this least-squares error approach to fitting a straight line over the unwrapped phase spectrum is equivalent to traditional cross-correlational TDE. Alternatively, it is possible to perform a linear fit over the wrapped phase spectrum by minimizing the squared residuals of Eq.~\eqref{eq:phasetodelay}. The delay estimate of both approaches can be further rounded to the nearest time bin. 

However, cross-spectrum based TDE is not restricted to the noiseless case. Similar to cross-correlational TDE, we can show that unmixed noise does not influence the cross-spectrum based TDE procedure. Consider again the segmented unmixed noise time series $X_s,Y_s$ with their respective Fourier coefficients $F_X$, $F_Y$. By nature of linearity of the Fourier transform, we can write $F_X(t)= \alpha F_x(t)+(1-\alpha)(F_{n_X}(t)+\theta_1 F_{n_Y})$ and $F_Y(t) = \alpha\beta F_y + (1-\alpha)(F_{n_Y}(t)+\theta_2F_{n_X}(t))$, and conclude that the resulting phase spectrum

\begin{align}
P_{XY,\text{unmixed}} = \angle \left(\alpha^2\beta \langle F_xF_y^* \rangle\right) \label{eq:punmixed2}
\end{align}

\noindent is free from contributions of noise for a sufficiently large number of segments (see Supplement, Eq.~\eqref{supp:phaseunmixed} for the full derivation). Note that the dependence of $P_{XY}$ and $F_{x,y}$ on frequency of is omitted here for simplicity. The slope of the phase spectrum is therefore invariant to unmixed noise influence, and TDE can still be expected to be reliable. 

The same cannot be said for cross-spectral phase-slope based TDE in mixed noise environments. Here, contributions from mixed noise generally compromise the amplitude and thus also the slope of the phase spectrum. This biases the overall estimate. Again omitting the dependence of $P_{XY}$ and $F_{x,y}$ on frequency, we can write 

\begin{align}
\nonumber P_{XY \text{, mixed}} = & \; \angle(\alpha^2\beta \langle F_xF_y^*\rangle \ldots\\
&+(1-\alpha)^2(\theta_2\langle\underbrace{|F_{n_X}|^2}_{\geq 0}\rangle+\theta_1\langle\underbrace{|F_{n_Y}|^2}_{\geq 0}\rangle) \;. 
\end{align}

\noindent The exact effect of mixed noise on the phase spectrum will depend on the sign of the noise term and is determined by $\theta_{1,2}$. For positive $\theta_{1,2}$, we can make use of the inequality 

\begin{align}
\left|\arctan2(Re+c,Im)\right| \leq \left|\arctan2(Re,Im)\right|,  \forall c \geq 0,
\end{align}
\noindent for the domain $[-\pi,\pi]$ to conclude that

\begin{align}
 \left|\angle\left( \langle F_xF_y^*\rangle+(\theta_2\langle|F_{n_X}|^2\rangle+\theta_1\langle|F_{n_Y}|^2\rangle\right)\right| \leq & \left|\angle\left( \langle F_xF_y^* \rangle \right)\right| \label{eq:pmixeduneq},
\end{align}
 \noindent or, 
 \begin{align}
 |P_{XY \text{, mixed}}|\leq & |P_{XY \text{, unmixed}}| \;.
\end{align}
 
\noindent Here, the scaling factors $\alpha^2\beta$ and $(1-\alpha)^2$ have been omitted. Depending on the size of the mixing effect $\theta$, as well as the absolute ($|F_{n_{x,y}}|^2$) and relative ($(1-\alpha)^2$) power of the noise components, the magnitude of the phase spectrum is thus compromised. With increasing noise contribution, the magnitude and slope of the phase spectrum approach zero. Similarly, one can show that for negative $\theta_{1,2}$, mixed noise contributions increase the magnitude of the phase spectrum, and can ultimately saturate the phase spectrum at $\pm \pi$. For unequally signed $\theta_{1,2}$, the net sign of the noise terms determines their effect on the phase spectrum, and will depend on the relative magnitudes of $\theta_{1,2}$ and $F_{n_{Y}},F_{n_{Y}}$. Overall, it is, thus, expected that the estimate $\tau_\text{phase-slope}$ deviates from the true delay $\tau$ in mixed noise environments.

\subsection{Bispectrum based TDE}

\noindent \cite{Nikias1988} proposed bispectrum (BS) based TDE approaches that are designed to suppress the influence of mixed noise. The bispectrum is the third-order moment generalisation of the cross-spectrum and is defined as 
\begin{align}\label{eq:bispectrum}
    B_{XYZ}(f_1,f_2) &= \langle F_X(f_1) F_Y(f_2) F_Z^*(f_1+f_2) \rangle,
\end{align}
for segmented signals $X_s,Y_s,Z_s$ and their respective Fourier transforms. The BS for a given combination of time series $X, Y$, and $Z$ is thus a two-dimensional quantity with dimensions $M_F\times M_F$. In TDE, $Z_s$ is replaced by either $X_s$ or $Y_s$ for the purpose of estimating delays between two channels. Higher-order moments evaluate to zero for Gaussian, or, more generally, zero-skewed sources  \cite{Nikias1993}, which can be exploited for data analysis in noisy settings. \cite{Nikias1988} specifically observe that bispectrum based TDE is possible under mixed noise influence for signals with nonzero skewness and Gaussian noise terms. This property is of particular relevance to brain data analysis, where the study of oscillatory and thus non-Gaussian components of electrophysiological data is driving current efforts in understanding cross-regional communication within the brain \cite{Buzsaki2011,Aru2015,Pellegrini2023} and higher-order cognitive functioning \cite{Chen2011}.

\cite{Nikias1988} present four TDE methods based on the BS. All make use of a bispectral hologram $h_{XY}$, which relates the cross-bispectrum $B_{XYX}$ to the auto-bispectrum $B_{XXX}$ (or $B_{YYY}$) and should, given an underlying time-delay $\tau$ between the two time series, have its maximum at $\tau$. Each of the four methods introduced by \cite{Nikias1988} define $h_{XY}$ slightly differently:

\begin{align}
    h_{XY}(\rho) &= IFFT\left(\sum_{f_2} I(f_1,f_2) \right) \label{eq:hologram},
\end{align}
    \noindent where
\begin{align}
    I_{M1}(f_1,f_2) &= \exp\left(i\left(\angle B_{XYX}(f_1,f_2)-\angle B_{XXX}(f_1,f_2)\right)\right) \\
\nonumber     I_{M2}(f_1,f_2) &= \exp\Bigg(i\Bigg(\angle B_{XYX}(f_1,f_2)\ldots \label{eq:hologram2}\\
    -\frac{1}{2}\left( \right. & \left. \angle B_{XXX}(f_1,f_2)+\angle B_{YYY}(f_1,f_2)\right)\Bigg)\Bigg)\\
    I_{M3}(f_1,f_2) &= \frac{B_{XYX}(f_1,f_2)}{B_{XXX}(f_1,f_2)}\\
    I_{M4}(f_1,f_2) &= \frac{|B_{XYX}(f_1,f_2)|\ I_{M2}(f_1,f_2)}{\sqrt{|B_{XXX}(f_1,f_2)|\ |B_{YYY}(f_1,f_2)|}} \;. \label{eq:hologram4}
\end{align}
\noindent From  $h_{XY}(\rho)$, a delay estimate $\tau_{\text{bispec.}}$ is derived:
    
\begin{align}
    \tau_{\text{bispec.}}=\arg\underset{\rho}{\max}\ h_{XY}(\rho) \;.
\end{align}
    
\noindent TDE in the bispectrum domain is comparable to cross-spectrum based TDE in that it aims to extract the phase difference of two signals. We demonstrate this for $I_{M1}$, for which, by virtue of Eq.~\eqref{eq:bispectrum}, the exponent reduces to

\begin{align} 
\nonumber \angle B_{XYX}&(f_1,f_2)-\angle B_{XXX}(f_1,f_2) \\ 
\nonumber =& \;\langle\varphi_{X}(f_1)+\varphi_{Y}(f_2)-\varphi_{X}(f_{1+2})\rangle \ldots\\
\nonumber & - \langle\varphi_{X}(f_1)+\varphi_{X}(f_2)-\varphi_{X}(f_1 + f_2)\rangle \\
=& \; \langle\varphi_{Y}(f_2)-\varphi_{X}(f_2)\rangle \label{eq:phasetodelay3}
\end{align}
\noindent for two given time series and a single frequency pair $(f_1, f_2)$. Comparing Eq.~\eqref{eq:phasetodelay3} to Eq.~\eqref{eq:phasespec} we see that \cite{Nikias1988} here effectively approximate the negative phase spectrum $-P_{XY}(f_2)$ with third-order moment terms. However, instead of estimating the slope of the phase spectrum, $I_{M1}$ constructs a panel of complex numbers which capture the periodicity and monotonicity of the phase spectrum. We note that $I_{M3}$ is similar to $I_{M1}$, with the addition of an amplitude-weighting of the contrasted bispectral frequency components. Similarly, it can be shown that the exponents of $I_{M2}$ and $I_{M4}$ equally reduce to 

\begin{align} 
   &\angle(I_{M2}) = \angle(I_{M4}) = \frac{1}{2} \langle\varphi_{X}(f_1)-\varphi_{Y}(f_1)\rangle \ldots \\
\nonumber      &- \frac{1}{2} \langle\varphi_{X}(f_2)-\varphi_{Y}(f_2)\rangle - \frac{1}{2} \langle\varphi_{X}(f_1+f_2)-\varphi_{Y}(f_1+f_2)\rangle,
\end{align} 

\noindent corresponding to a combination of phase spectra of the different frequency components (for the full derivation, see Eq. \eqref{eq:IM2,IM4phase2} in the Supplement). Note also that $I_{M4}$ corresponds to an amplitude-weighted $I_{M2}$.

In case of bispectrum based TDE, it becomes more difficult to analytically disentangle the contributions of signal and noise to the final delay estimate. It is, however, useful to understand their combined effect on the BS itself. Assuming independent signal and noise, we can expand $F_X(f) = F_x(f) + F_{n_X}(f)$ (and similarly for $F_Y,F_Z$), and substitute the result into Eq.~\eqref{eq:bispectrum} to show that the combined BS of signal and noise corresponds to a superposition of their individual bispectra: 

\begin{align}
\nonumber B_{XYZ}(f_1,f_2) &= \langle F_X(f_1) F_Y(f_2) F_Z^*(f_1+f_2) \rangle \\
& = \left\langle F_x(f_1)F_y(f_2)F_z^*(f_1+f_2) \right\rangle \dots \label{eq:combinedbispec}\\
\nonumber &+ \left\langle F_{n_X}(f_1)F_{n_Y}(f_2)F_{n_Z}^*(f_1+f_2) \right\rangle + c.t. \;.
\end{align}

\noindent Here, ($c.t.$) denotes coupling terms that contain products of signal and noise variables whose expectations evaluate to zero. In the noiseless case as well as when the noise sources are uncorrelated, the second term in \eqref{eq:combinedbispec} also equals zero. Moreover, the second term vanishes for mixed noise sources if they are Gaussian distributed \cite{Isserlis1918}. In practice, Gaussian suppression of noise contributions has been observed to work more or less well, depending on the algorithm used to approximate the BS \cite{Nikias1988}.  The reason for this is still unclear. More generally though, current BS based TDE methods are conceived to perform well in all of the above considered settings. The performance of bispectral TDE in other settings is yet to be assessed. 

Evaluating the effects of non-Gaussian mixed noise on bispectral TDE is less straightforward. Here, the noise term in Eq.~\eqref{eq:combinedbispec} is generally nonzero and the resulting BS will contain contributions from both signal and noise components. The phase of the combined BS is a nonlinear function of the phases of the individual components, and it is in turn not expected that the bispectral hologram will consistently have its maximum at an underlying delay $\tau$. How bispectral TDE is influenced by mixed noise precisely is yet to be outlined. 

\subsection{Robust TDE using Phase Periodicity}
\noindent In Section~\ref{sec:cross spec TDE}, we discuss that the phase-slope delay estimate is biased towards zero (or $\pm \pi$, depending on the polarity of the signal) in mixed noise environments, where the magnitude of the phase spectrum is compromised. Instead of the slope of the phase spectrum, we therefore propose to estimate its periodicity, which remains unaffected by mixed noise. 

The periodicity of a sawtooth function is a property as intrinsic as its slope. For time-delayed sources, one period of the phase difference spectrum is defined by the points of phase wrapping, i.e., where the sign of the phase spectrum flips at $\pm \pi$. At constant slope, phase wrapping occurs periodically, and the periodicity of the phase spectrum is equally proportional to the underlying delay. It is therefore, in theory, a suitable candidate for a cross-spectral delay estimator. 

In fact, phase periodicity may be a more stable delay estimator than the slope of the phase spectrum. In the noiseless case, we consider both approaches to be equally suitable. The same holds for the unmixed noise case, for which we have shown that noise contributions to the phase spectrum cancel out in the expectation (see Equation \eqref{eq:punmixed2}). However, while mixed noise diminishes the slope of the phase spectrum, its periodicity is expected to be conserved. The effect of mixed noise on a periodicity based delay estimate should, therefore, be restricted to a decrease in statistical power of the analysis, but not introduce a bias. 

It is important to note that, with a periodicity based cross-spectral TDE approach, information about the directionality of the delay is lost. It therefore becomes necessary to estimate this directionality with another metric. In our proposed TDE approach, we determine the directionality of the delayed interaction with the sign of the Phase Slope Index (PSI, $\Psi$) \cite{Nolte2008}. Given a set of frequencies $F$ with a frequency resolution of $\delta f$, the (unnormalized) PSI is defined as

\begin{align} \label{eq:PSI}
\Psi = \Im\left(\sum_f^{F}CHY_{XY}(f)\times CHY^*_{XY}(f+\delta f)\right),
\end{align}

\noindent where $CHY_{XY} = {S_{XY}}/{(S_{XX}S_{YY})^{1/2}}$ is the complex-valued coherency and $\Im(\cdot)$ denotes taking the imaginary part. The PSI has been shown to correspond to a weighted average of the slope of the phase spectrum \cite{Nolte2008} and is, importantly, also insensitive to mixing artifacts of combined independent sources \cite{Nolte2004}. The sign of the PSI can thus serve as a metric for the directionality of the flow of information between two channels that is robust to mixed noise. This information is then used together with an estimate for the periodicity of the phase spectrum to approximate the value of an underlying delay:

\begin{align}
\nonumber \tau_{\text{period.}} &= \text{sign}(\Psi)\ \arg\max_t\ \left|FFT\Big(P_{XY}(f)\Big)\right| \\
&= \text{sign}(\Psi)\ \arg\max_t\ \left| \sum_{f=0}^{T-1} P_{XY}(f)\ e^{-\frac{i2\pi}{T}ft}\right| \;.
\end{align}

\subsection{Robust Bispectrum based TDE using Antisymmetrization}
\noindent In the BS domain, mixing artifacts can be addressed by means of antisymmetrization \cite{Chella2014}. The antisymmetrized BS (ASB) is defined as the the difference between the cross-bispectrum $B_{XYZ}(f_1,f_2)$ and a permutation of it with respect to any pair of channel indices. In their original paper, \cite{Chella2014} show that the ASB cannot originate from independent sources by demonstrating that it vanishes for any linear combination of independent sources. A nonzero ASB thus captures nonlinear source interactions, and is free from mixed noise contributions. We therefore propose to adopt antisymmetrization into the bispectral TDE framework by replacing the cross-bispectrum $B_{XYX}$ in Eq.~\eqref{eq:hologram} with its antisymmetrization
\begin{align}
B_{[X|YX]}:= B_{XYX}-B_{YXX} \;. 
\end{align}

\noindent We will refer to the resulting antisymmetrized bispectral delay estimate as $\tau_{\text{ASB}}$.

\subsection{Confidence based TDE} \label{sec:conf}

\noindent All conventional and novel TDE approaches introduced here are forced-choice. This means that even when there is no distinct interaction between a pair of time series, all TDE approaches will output a delay estimate. It therefore becomes necessary to discern reliable from unreliable delay reports. For this purpose, we derive a confidence-based statistical test that can be combined with standard TDE procedures.

We derive a confidence filter applicable to every TDE method. 
based on the null hypothesis that the underlying interaction between $X$ and $Y$ is either non-existent or due to a zero-valued delay. The alternative hypothesis is that the underlying delay is nonzero. 
If, for a given delay estimate, we observe zero to be outside the CI bounds, we reject the null and accept the alternative hypothesis.

Testing whether the resulting estimate is significantly different from zero ensures two things. First, it rules out that the estimate reflects instantaneous interactions possibly arising from mixed noise. Second, our confidence filter recognizes low confidence reports which may arise when no interaction between $X$ and $Y$ is present at all. In this case, both the phase spectrum and the bispectrum are undefined. 

\section{Simulations} 
\label{sec:expGauss}
The present study contains three experimental sections: in the present section, we contrast the outcomes of TDE approaches for temporally uncorrelated (white) Gaussian and non-Gaussian noise environments. We also provide a qualitative analysis of the cross-correlogram, the phase spectrum, and the bispectral hologram for each noise setting. In Section \ref{sec:expreal} (\textit{Electrophysiological Data}), we test the efficacy of our TDE approaches on a small electrophysiological dataset where signal time-delays are to be expected. Finally, in Supplementary Section~\ref{sec:expcolor} (\textit{Simulations with Auto-correlated Noise}), we repeat analyses of Section~\ref{sec:expGauss} for auto-correlated non-Gaussian noise sources. 
All analyses were performed in MATLAB\textregistered\ R2020b (9.9.0. 1467703). The source code for this project will be made openly available upon publication.


\subsection{Methods}
\noindent \subsubsection{Data Generation}
All TDE methods are tested on an artificial two-channel dataset. The time series of each channel has a length of $t = 130$ seconds at a sampling rate of $f_s = 100$ Hz and consists of a signal and a noise component. One time bin thus corresponds to an interval of 10 milliseconds (ms). Since we aim to resolve differences in TDE between Gaussian and non-Gaussian noises, we sample the noise either from a standard normal ($\mathcal{N}(0,1)$) or an exponential ($Exp(\lambda=1)$) distribution. The signal component is drawn from the same exponential distribution, but independent of the noise. The observed time series are then assembled according to Eq.~\eqref{eq:fullTDE}, where we choose $\beta =1$ and $\theta_{1,2}=.7$. Time delays estimated from this mixed-noise dataset are compared to those estimated from a dataset with unmixed noise, which is obtained by setting $\theta_{1,2} = 0$.

For each signal and noise combination, we generate $n_{\text{trial}} = 500$ channel pair time series (trials) with random, non-zero delays within the interval bounds $[-100,100]$. Each trial is assembled at varying SNR $\alpha = [0,1]$ in increments of $0.1$ (note that we normalize both signal and noise times series by their Frobenius-norm prior to summation, so the relation between $\alpha$ and decibel SNR is given by $SNR_{dB} = 10 \log_{10}\frac{\alpha}{1-\alpha}$). Delays are estimated via conventional and novel cross- and bispectrum based TDE approaches as introduced above. 

\subsubsection{Application of TDE approaches}
\noindent Before applying TDE approaches, the data are segmented into $n_{\text{segment}} = 65$ non-overlapping segments of $l_{\text{segment}}=2$ seconds. We here restrict ourselves to five TDE approaches, three conventional ones known to be negatively affected by mixed noise and two novel methods conceived to overcome that limitation. Concretely, we include: (1) cross-correlational TDE (2) cross-spectrum based TDE implemented by estimating the slope of the phase spectrum, (3) the proposed novel cross-spectrum based TDE implemented by estimating the periodicity of the phase spectrum combined with the directionality of the delay estimated by $\text{sign}(\Psi)$, (4) BS based TDE as described in \cite{Nikias1988}, and (5) the proposed novel ASB TDE approach combined with ASB \cite{Chella2014}. Out of the four BS TDE methods presented in \cite{Nikias1988}, we include only \textit{Method MI} (Eq.~\eqref{eq:hologram}) in our analyses, as it appeared to be the most numerically stable in preliminary experiments (for a comparison of antisymmetrized bispectral holograms derived via the four methods, see Supplementary Figure~\ref{supp:m14}). 

\subsubsection{Confidence of Delay Estimates} \label{sec:mconf}
\noindent For each cross- and bispectrum based delay estimate, we apply a respective confidence filter as outlined in Section~\ref{sec:conf}. We sample $n_{\text{segment}} = N = 65$ segments with replacement for a total of $n_{\text{boot}} = 500$ iterations and perform TDE for each iteration separately. Trials that do not meet the confidence criterion are left out from further analysis. We first assess the behaviour of cross- and bispectrum based TDE filters: For each signal and noise environment, we determine the rate of rejected trials at increasing SNR. Datasets with SNR $\alpha = 0$ (with no signal contribution) are used to determine the specificity or false alarm rate of our filtering approaches, while its sensitivity or acceptance rate is assessed on datasets with non-zero $\alpha$.

\subsubsection{Quantifying Delay Estimation Errors}
\noindent For each trial, we obtain the absolute error (Eq.~\eqref{eq:err}) between true delay and average estimated delay:

\begin{align} \label{eq:err}
\text{AE} = |\tau_{\text{true}}-\overline{\tau_{\text{boot}}}| \;,
\end{align}
where the average $\overline{\tau_{\text{boot}}}$ is taken over bootstrap iterations.
The absolute errors are then averaged across trials for each SNR and each combination of signal and noise components to obtain a mean absolute error (MAE). We ultimately compare the MAE of trials that have been filtered according to our confidence criterion to those of the unfiltered dataset. We also investigate differences in accuracy across TDE approaches. 

\subsubsection{Identifying Biases in TDE}
\noindent To examine systematic delay estimation errors, we generate ten independent datasets for each delay $\tau_{\text{true}} = [-200,200]$ according to the procedure outlined above and plot estimated delays $\tau_{\text{est}}$ as functions of true delays. We further report Pearson's correlation coefficient (PCC) for datapoints with $\tau_{\text{true}}$ within [-100,100], corresponding to the segment length of the times series.

\subsection{Results}
\subsubsection{Qualitative Assessment}
\begin{figure}[htbp]
\centering
\includegraphics[width=\columnwidth]{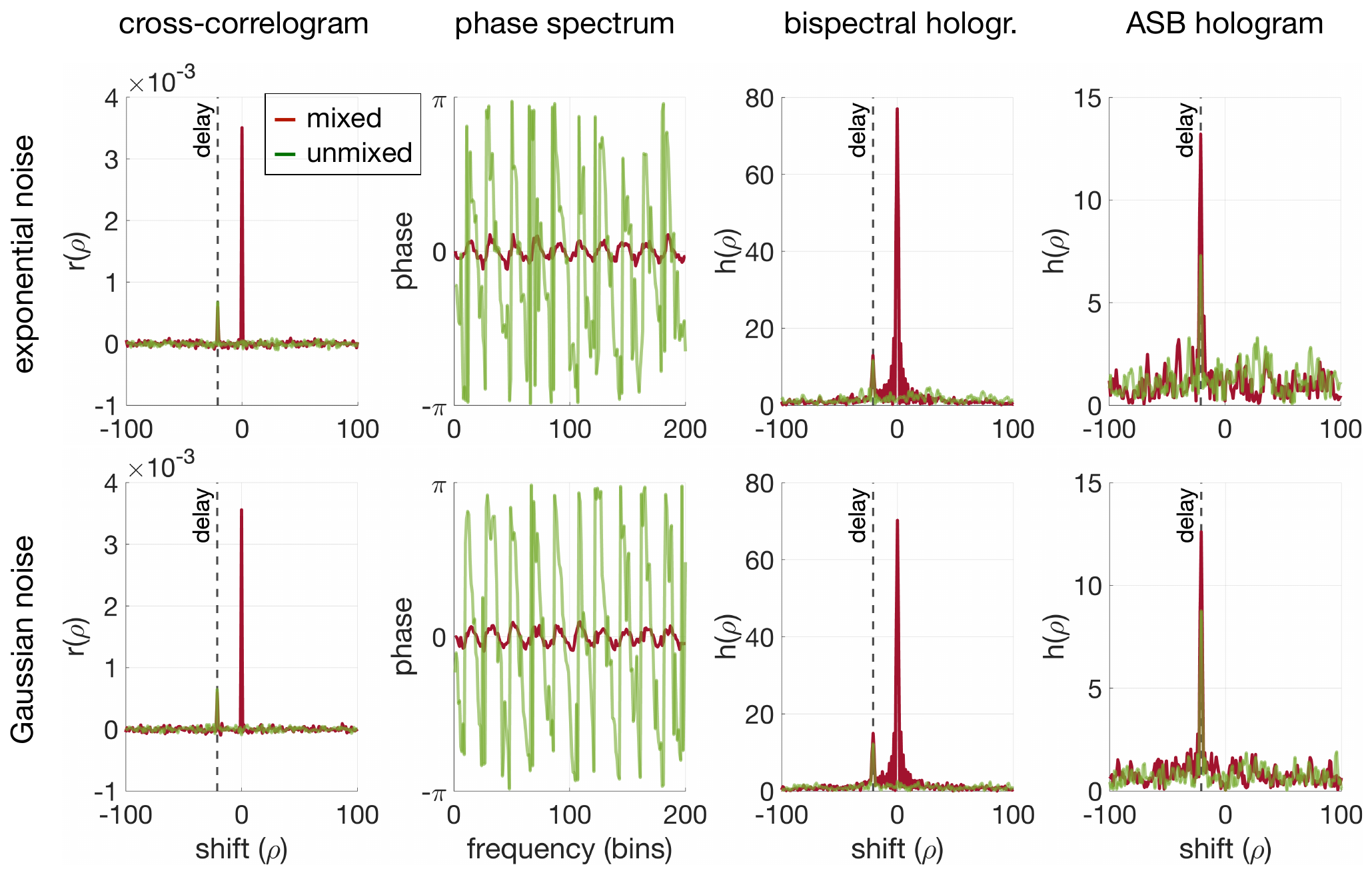}
\caption{Cross-correlogram, phase spectrum, bispectral hologram, and anti-symmetrized bispectral (ASB) for exponentially distributed (non-Gaussian) signal and exponentially distributed (top row) as well as Gaussian (bottom row) noise at SNR $\alpha = 0.4$. Mixed noise (red) contributions in both settings are reflected by zero-lag peaks in the cross-correlogram and bispectral hologram as well as a reduction in amplitude of the phase spectrum compared to the unmixed noise setting (green). Bispectral antisymmetrization removes spurious zero-lag peaks introduced by noise mixing.}
\label{fig:overviewG}
\end{figure}

\noindent We first visualize the cross-correlograms, phase spectra and bispectral holograms for exponential and Gaussian noise combinations. An example for SNR $\alpha = .4$ is provided in Figure~\ref{fig:overviewG}. For unmixed noise environments (green), we observe the cross-correlogram to be generally flat with a peak at the value of the underlying delay. For mixed noise environments (red), an additional peak at lag zero is introduced. All phase difference spectra show a characteristic sawtooth shape for sufficiently high SNR. In mixed noise environments, we observe the expected corruption of amplitude as early as for $\alpha = .8$ (not shown). Notably, the periodicity of the phase spectrum is conserved in all mixed noise environments. These results are generalizable across Gaussian and exponential (non-Gaussian) noise environments. 

Noise mixing introduces a zero-lag peak in the bispectral holograms for both exponential and Gaussian  noise sources, in addition to the peak of the underlying delay (Figure~\ref{fig:overviewG}). This stands in contrast to the theory laid out in \cite{Nikias1988}, which states mixed Gaussian noise should be suppressed by BS-based TDE approaches. Still, antisymmetrization appears successful in removing these mixed noise contributions, both in the exponential and Gaussian noise case. Bispectral TDE in combination with antisymmetrization thus appears to be a viable candidate for TDE in mixed noise environments.

\subsubsection{Quantifying TDE Accuracy}

\begin{figure*}[htbp]
\centering
\includegraphics[width=0.8\textwidth]{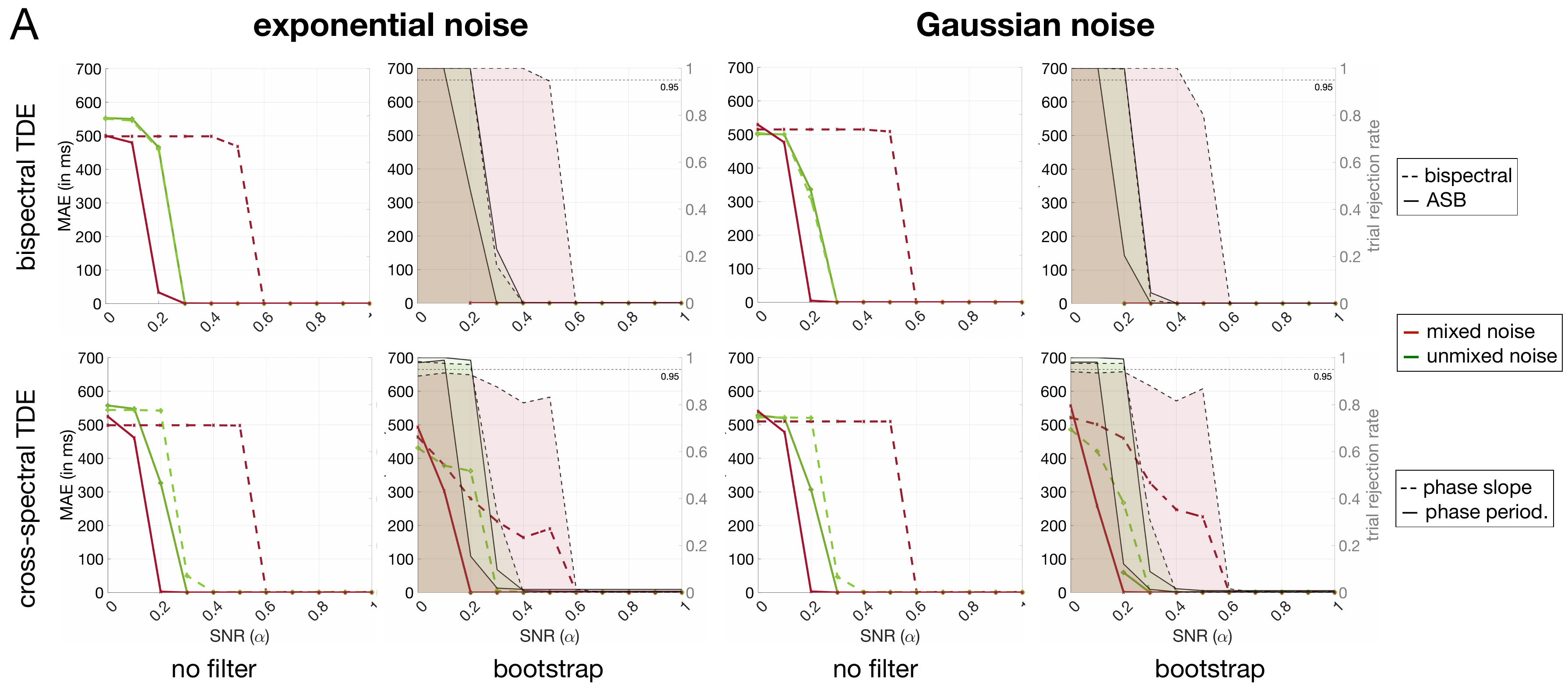}
\includegraphics[width=\textwidth]{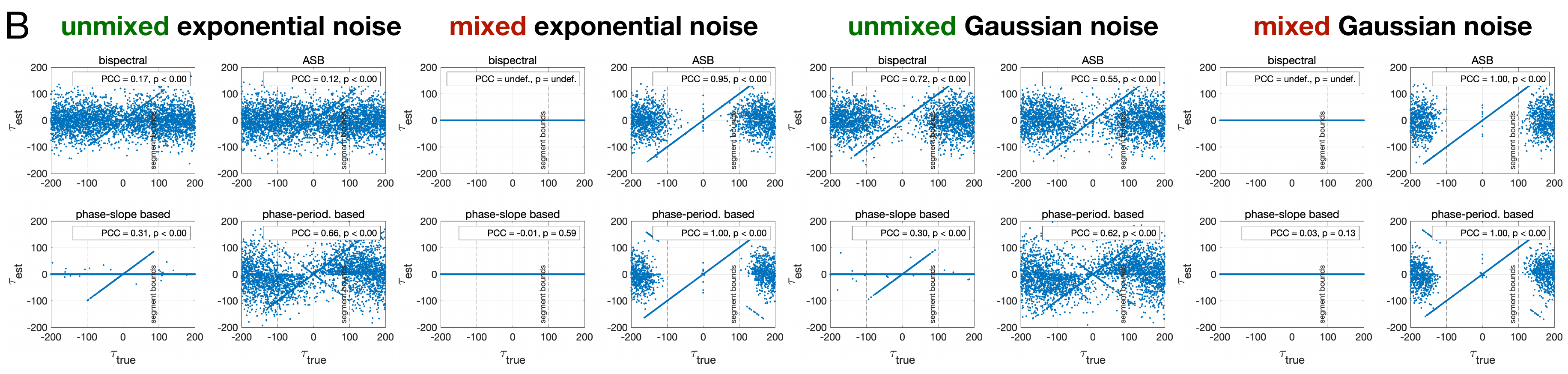}
\caption{Performance of different time delay estimation (TDE) approaches. A: Performance of cross-spectrum (bottom row) and bispectrum (top row) based TDE approaches in exponential (left panel) and Gaussian (right panel) noise environments. Depicted are the mean absolute errors across $n_{\text{trials}} = 500$ trials for each SNR $\alpha = [0,1]$. The results obtained for each noise environment are presented with either no filter or a bootstrap-based confidence filter applied. The shaded areas indicate the rejection rate for each set of trials. Red color marks the results for mixed noise trials, while green color is used for results of unmixed noise trials. Conventional bispectrum and phase-slope based TDE approaches (dashed lines) fail in the presence of mixed noise for low SNR, whereas our antisymmetrized bispectrum and phase-periodicity approaches remain sensitive to signal interactions in mixed noise environments for SNRs as low as $\alpha = .2$. The proposed confidence filter led to successful rejection of the most incorrect estimates, driving the overall mean absolute error to near zero in almost all cases. B: Distribution of true ($\tau_\text{true}$) versus estimated ($\tau_\text{est}$) delays at SNR $\alpha$ = 0.2 (n=10 trials per $\tau_\text{true} = [-200,200]$). 
A high fraction of correct delay estimates is visible in unmixed noise environments as a diagonal line. For ASB and phase-periodicity TDE approaches, the same is also discernible in mixed noise environments, whereas conventional bispectrum and phase-slope TDE methods tend to report spurious zero-delays in presence of mixed noise. The horizontal lines indicate the segment bounds of the generated dataset. Pearson's correlation coefficient (PCC) and corresponding p-values are reported for true lags contained within the segment bounds $[-100;100]$.}
\label{fig:perfG}
\end{figure*}

\noindent We first inspected the MAE for all TDE approaches with no filter applied. All results are summarized in Figure~\ref{fig:perfG}A. For bispectrum-based TDE approaches, $MAE \approx 500$ ms for SNR $\alpha = 0$, i.e. where no signal contributed to the combined timeseries (see Figure~\ref{fig:perfG}A). In the unmixed noise setting (green lines), conventional BS (dashed line) and ASB TDE approaches perform similarly, with their respective MAEs decreasing to near zero at SNR $\alpha =.3$. In the mixed noise setting (red lines), ASB has a clear advantage over its unsymmetrized counterpart: whereas the MAE remained at around $500$ ms for conventional bispectral TDE until SNR $\alpha = .5$, the MAE decreases to near zero as early as for $\alpha = .2$ for the antisymmetrized counterpart, thus even improving upon its result achieved in the unmixed noise setting.. 
These observations are made both for Gaussian and exponentially distributed non-Gaussiannoise environments. 

The findings for cross-spectrum based TDE, are comparable: for unmixed noise settings and and both cross-spectrum based approaches, the MAE is near zero for $\alpha \geq .4$ (Figures~\ref{fig:perfG}A). 
We observe a slightly lower MAE when applying our phase-periodicity based (full line) approach compared to its phase-slope (dashed line) counterpart. In the mixed-noise setting, differences in performance between the two methods are more pronounced. For phase-periodicity based TDE, the MAE is near zero for $\alpha \geq .2$, again even improving upon its result in the unmixed noise setting, whereas phase-slope based TDE does not show any signal sensitivity for SNRs lower than $\alpha = .5$, keeping the MAE comparable to the setting with no signal present ($\alpha = 0$).

\subsubsection{Confidence Filters}

\noindent Figure~\ref{fig:perfG}A also shows the same results as previously discussed but with the bootstrapped confidence filter applied. The rejection rates for each SNR and method are indicated by the shaded areas in red and green for mixed and unmixed environments, respectively. The rejection rate for $\alpha = 0$ is 1 in all four panels, indicating a false alarm rate of 0 for all filtering approaches. The width of the confidence interval was chosen to be $w = .95$ for all four panels. When increasing SNR, we consistently observe the exclusion rate to decrease at around $\alpha = .3$, dropping to near zero above this value. An exception are conventional TDE approaches in the mixed-noise setting (red filling, dashed outline), where the exclusion rates do not drop until SNR $\alpha = .6$. 
Confidence filtering appears to be a feasible approach to improving TDE accuracy, as we observe it to drive the mean absolute error to zero in almost all settings. Minor exceptions are observed for conventional phase-slope based TDE in both exponential 
and Gaussian noise environments (Figure~\ref{fig:perfG}A), where few trials accounted for non-zero MAEs in lower SNR regimes ($\alpha < .6$). 

Taken together, we observe similar behaviour of cross- and bispectrum based TDE approaches in exponential and Gaussian noise environments. Our novel phase-periodicity and antisymmetrized bispectrum based TDE approaches successfully circumvent estimation errors related to mixed noise contributions. Combined with a bootstrap-based confidence filter, they achieve mean absolute errors close to zero. 

\subsubsection{Distribution of TDE outputs}

\noindent With increasing SNR, we observe TDE errors to approach zero. Here, the delay estimate is equal to the underlying true delay. In regimes where the MAE is not equal to zero, TDE outcome might be ambiguous. To look into potential biases of TDE approaches, we study the cases of no signal and low SNR, $\alpha = 0$ and $\alpha = 0.2$. These distributions are summarized in the bottom panel of Figure~\ref{fig:perfG}.

In the case where no signal is present at all (SNR $\alpha = 0$), TDE outputs of all approaches are evenly distributed (top two rows). In the unmixed noise case, both bispectrum-based approaches and the phase-periodicity based TDE approach produced random outputs, whereas the phase-slope approach consistently estimates a delay of zero. In the mixed noise case, conventional bispectrum approaches also output zero-delay estimates across the entire sampled range. PCC was either near-zero, zero, or undefined for all settings ($P \geq .14$). These results are independent of the underlying noise distribution. 

At nonzero SNR, $\alpha = 0.2$ (bottom two rows), we observe a linear true-to-estimated delay behavior of TDE approaches in unmixed noise environments for delays within segment bounds $[-100;100]$. A similar relationship between input and outputs for phase-periodicity and ASB based TDE approaches was found in mixed noise settings. For all these settings, we report nonzero PCC with $p < 0.00$. For conventional TDE approaches, the estimated delays in mixed noise settings again are almost always zero. Again, these results are independent of the underlying distribution of the noise component. For higher SNR, $\alpha \geq 0.3$, PCC steadily increases to 1 (not shown).

\section{Electrophysiological Data}\label{sec:expreal}
\noindent In a second experiment, we explore the viability of the proposed TDE approaches on electrophysiological data. 

Electrophysiology concerns itself with the measurement of electrical potential differences reflecting the effect of summed electrical activity of underlying neurons. Event-related electroecephalography (EEG), for example, measures the electrical brain response to an external stimulus. Repeated over several dozen or hundred times, an averaged event-related potential (ERP) can be obtained, and parameters of interest (e.g. peak shapes and latencies) can be extracted. 

ERPs of the somatosensory cortex are well documented in response to peripheral stimulation and pose clinical relevance \cite{allison1991potentials,cruccu2008recommendations}. Stimulating the median nerve at the wrist of the hand, the ERP exhibits a characteristic negative deflection at 20~ms after stimulus onset, the so-called N20 component, visible at electrode positions posterior of the central sulcus. The somatosensory cortex has its afferents in the contralateral periphery, and the amplitude of the N20 deflection has been linked to the somatosensory cortex' excitability to external stimuli at a given time \cite{stephani2020temporal}. Delays in the N20 amplitude have further shown to be relevant in neurorodegenerative diseases such as amyotrophic lateral sclerosis \cite{Zhang2014}. More generally have neuronal delays been suggested as a useful metric to investigate signal propagation and cortical hierarchies across different signaling pathways \cite{Stam2007}.

Currently, ERP analysis requires prior knowledge of the stimulus timing in order to perform time-locking and obtain an averaged response. In the context of signal latencies, our novel TDE approaches thus pose a promising alternative to ERP analysis, as they are able to (i) extract delays values between channel pairs without prior information about stimulus timing and (ii) are robust against mixing artifacts common to EEG analysis. On a subset of peripheral stimulation EEG data assessed by \cite{stephani2020temporal}, we here want to compare latencies between peripheral and cortical electrodes by means of ERP and TDE analysis, and discuss our findings with respect to the applicability of our TDE approaches on neural data. 

\subsection{Methods}
\subsubsection{Experimental Setup}
\noindent For a detailed description of the considered experimental dataset, the reader is referred to \cite{stephani2020temporal}. Briefly, the experiment comprised median nerve stimulation of $31$ healthy adult human participants at the left wrist with the following concurrent neural measures: assessment of the excitation of the median nerve in the periphery (measured as compound nerve action potential (CNAP) at the level of the ipsilateral upper arm using a pair of bipolar electrodes), as well as recording of the cortical response from 60 EEG electrodes (measured with an 80-channel EEG system; NeurOne, Bittium, Oulu, Finland).
The electrical stimulus was designed as a 200 $\mu s$ squared pulse and presented a total of 1,000 times with interstimulus intervals (ISI) ranging from 663 to 763~ms. Peripheral and EEG responses were captured at a sampling rate of 5~kHz.

\subsubsection{Preprocessing}
\noindent Preprocessing of the EEG data included stimulation artifact interpolation, bandpass-filtering at [30, 200] Hz, and re-referencing to the average of all channels. Eye movement artifacts were removed using independent component analysis \cite{delorme2004eeglab}. CNAP recordings were highpass-filtered at 70 Hz and additionally notch-filtered at [48, 52] and [148, 152] Hz.

\subsubsection{TDE Analysis}
We limit the present analysis to three subjects, S12, S17, and S27, based on the goodness of raw data quality. Prior to application of TDE approaches, the data are downsampled to 1,250~Hz and segmented into intervals of 200~ms duration starting 15~ms before stimulus onset. This way, we generate 1,000 segments per subject, each containing 250 time bins, with a spacing of $dt=0.8$~ms. In a subsequent analysis, segmentation is not time-locked to the stimulus onset. Here we again segment the downsampled data into 200 ms intervals, but this time across the entire time-series, amounting to 3628 segments.

We then apply our antisymmetrized bispectrum and phase periodicity based TDE approaches between the CNAP channel and three EEG channels placed above the right somatosensory cortex, F4, CP4, and P4. We adopt a similar bootstrapping procedure as outlined in Experiment I (Section~\ref{sec:expGauss}) with $n_{\text{boot}} = 500$ to determine the confidence of TDE outputs. Instead of filtering the results, we plot the entire distribution of obtained delay estimates and report its median. Ultimately, we compare the outcomes of stimulus-locked and non-time-locked analyses.

\subsection{Results} 

\begin{figure}[htbp]
\centering
\includegraphics[width=\columnwidth]{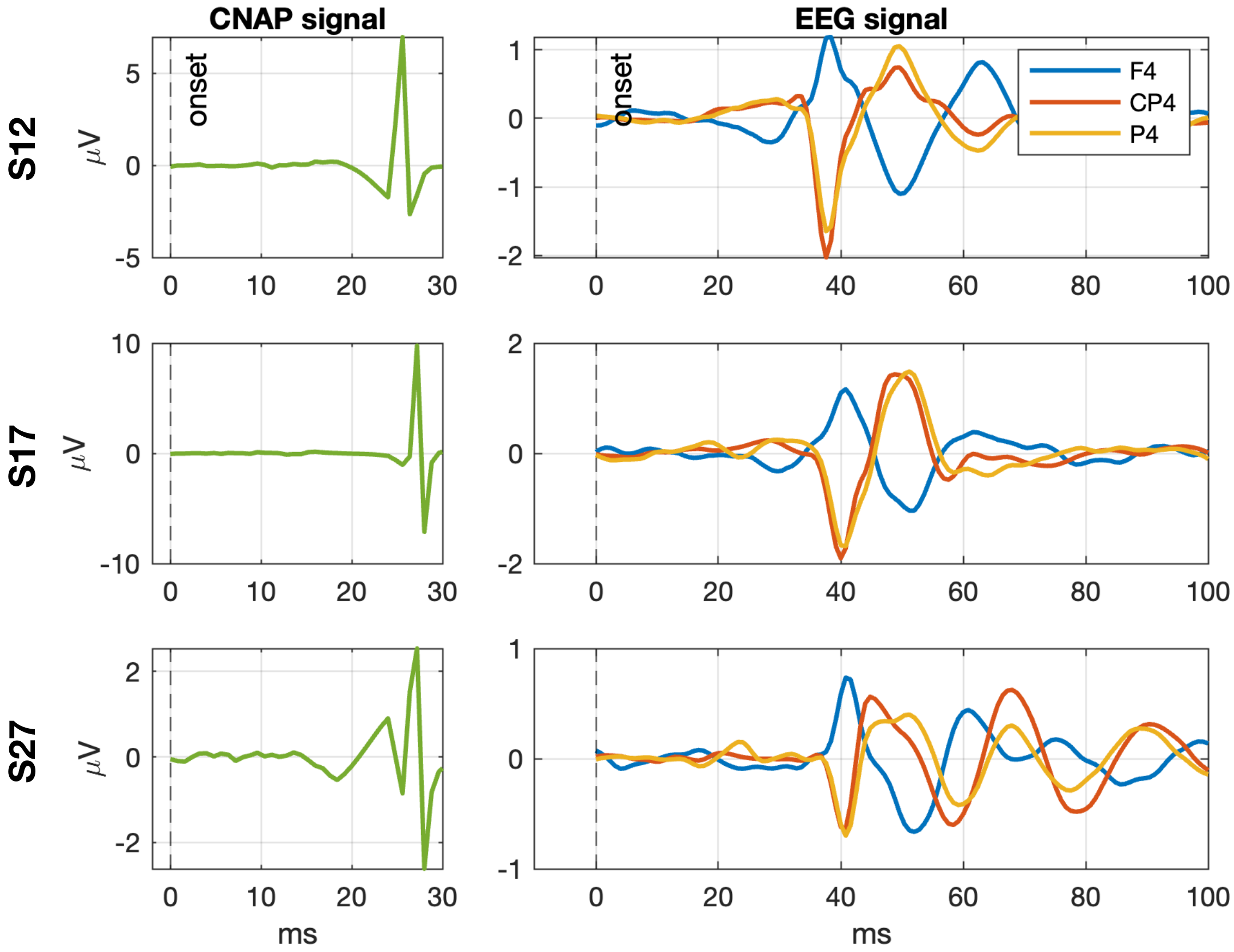}
\caption{Averaged response of peripheral (CNAP) and cortical (EEG) electrodes in subjects S12, S17, and S27. }
\label{fig:erps}
\end{figure}

\subsubsection{Averaged Responses}

Figure~\ref{fig:erps} shows the averaged responses of CNAP and selected somatosensory EEG electrodes for all three subjects. The overall average response amplitude was highest for subject S17, where the CNAP response ranged between $[-7, 10]$~$\mu$V and EEG responses were bounded by $[-2, 2]$~$\mu$V. For subjects S12 and S27, the CNAP response ranged between $[-3, 7]$~$\mu$V and $[-2, 2]$~$\mu$ V, and the EEG responses bounded by $[-2, 1]$~$\mu$V and $[-1, 1]$~$\mu$V, respectively. 

The CNAP responses comprised an initial sink at $t=4-6$~ms and a sharp positive deflection at $t=6-8$~ms, succeeded by an immediate sharp negative deflection. In S27, we observe an additional peak shortly after stimulus onset, which may be a residual stimulation artifact. In the EEG channels of all subjects, we could identify the characteristic N20 amplitude in the CP4 and P4 electrodes, and a subsequent positive deflection at $\approx 30$ ms. We found a similar response of the F4 electrode, but with opposite polarity, which is expected based on the electrode´s position anterior to the central sulcus. The effective latencies between the positive deflection in the CNAP signal and the cortical N20 and 30~ms amplitudes were thus $\approx 13$~ms and $\approx 23$~ms, respectively. Whereas in subjects S12 and S17 the EEG responses were mostly limited to these two deflections, the EEG responses in S27 showed several more subsequent peaks. 

\subsubsection{TDE Analysis}
The medians and corresponding interquartile ranges of the obtained bootstrapped distributions are summarized in Table~\ref{tab:realdatatde}. The distributions of ASB and phase-periodicity based TDE outputs themselves are shown in Figure~\ref{fig:realdatatde} in the supplement. 
The analysis was performed twice, once on data segments time-locked to the stimulus onset (left panel of Figure~\ref{fig:realdatatde}), and once where segmentation was arbitrary (right panel). 

For subject S12 and stimulus-locked segments, we found that the medians of estimates consistently range between 10.4 and 12.8~ms (corresponding to three time bins) for all three electrode pairings and both ASB and phase-periodicity approaches. For electrode pairs CNAP$\to$CP4 and CNAP$\to$P4, these estimates satisfy our confidence criterion with a chosen CI width of 95\% (highlighted in blue). The median estimates for CP4 and P4 differ slightly between the two TDE approaches. For both channels, we report an estimated delay of 10.4~ms for ASB TDE and of 12~ms and 12.8~ms for phase-periodicity based TDE. While the median estimated delay of the peripheral and F4 electrode pairing is of the same order of magnitude as the other two, the estimate does not pass our confidence criterion.

On the arbitrarily segmented data of subject S12, we were able to reproduce the findings reported for stimulus-locked data for the CNAP$\to$CP4 electrode pairing for both the ASB and phase-periodicity based TDE approach. The median ASB delay estimates for the P4 channel are also identical to that of the stimulus-locked analysis. For the arbitrarily segmented CNAP$\to$F4 electrode pairing, ASB and phase-periodicity TDE outputs have identical medians of 7.2~ms, which are noticeably lower than the stimulus-locked pendants. However, none of the F4 TDE analyses in S12 passes the confidence criterion. 

For subjects S17 and S27, the distributions of bootstrapped delay estimates appear generally broader than those of S12, with median estimates ranging from -9.6 to 46.4~ms. Only few estimates in S17 pass our confidence filter: both ASB based and phase-periodicity based TDE of the stimulus-locked P4 electrode pairing, and the ASB based analysis of the CP4 electrode pairing with arbitrary segmentation. Here we estimated median latencies of 12,13.6, and 12.8~ms, respectively. These numbers are comparable to the corresponding analyses in subject S12 (all of which passed our confidence filter). For subject S27, TDE analysis with our cross- and bispectrum based approaches did not lead to well-localized delay estimates.

In all accepted trials, the TD estimates range between 10.4 and 12.8~ms. These values are close to the latencies between the CNAP response peak and the N20 peak recorded at the cortical electrodes in an ERP analysis ($\approx 13$ ms). Therefore, they appear physiologically plausible.

\begin{table*}[htbp]
\caption{Antisymmetrized bispectrum (ASB) and phase-periodicity based time delay estimates (TDE) on electrophysiological data. Values are given as median (interquartile range) as estimated from 500 bootstrap iterations  in units of milliseconds.}
\centering
\resizebox{0.7\textwidth}{!}{
\begin{tabular}{llllp{0mm}lll}
    \multicolumn{8}{l}{\textbf{Stimulus-locked Segmentation}} \\ 
    \toprule
     & \multicolumn{3}{l}{ASB TDE} & & \multicolumn{3}{l}{Phase-period. TDE} \\    
    \cmidrule{2-4} \cmidrule{6-8}
    & F4 & CP4 & \multicolumn{1}{l}{P4} & & F4 & CP4 & P4 \\
\cmidrule{2-4} \cmidrule{6-8}
S12 & 11.2 (3.2) & \textbf{10.4} (0.8) & \multicolumn{1}{l}{\textbf{10.4} (0.8)} & & 12 (24.4)   & \textbf{12} (0.8) & \textbf{12.8} (1.6) \\
S17 & 30.8 (52.8)         & 14.4 (44)                          & \multicolumn{1}{l}{\textbf{12} (0.8)} &  & 22.4 (46.8) & 18.4 (44.8)                      & \textbf{13.6} (0.8) \\
S27 & {8} (107.6)           & {42.4} (105.6)                       & \multicolumn{1}{l}{{12.8} (116.4)}  &                     & {-5.6} (39.2) & {24.8} (71.2)                      & {42.4} (60.4)                  \\   
\bottomrule
\\
\\
    \multicolumn{8}{l}{\textbf{Arbitrary Segmentation}} \\ 
    \toprule
     & \multicolumn{3}{l}{ASB TDE} & & \multicolumn{3}{l}{Phase-period. TDE} \\    
    \cmidrule{2-4} \cmidrule{6-8}
    & F4 & CP4 & \multicolumn{1}{l}{P4} & & F4 & CP4 & P4 \\
\cmidrule{2-4} \cmidrule{6-8}
S12 & {7.2} (20.4)   & \textbf{10.4} (0.8)  & \multicolumn{1}{l}{\textbf{10.4} (1.6)} & & {7.2} (54)    & \textbf{12} (1.6) & {11.2} (29.6) \\
S17 & {12.8} (126.8) & \textbf{12.8} (23.2) & \multicolumn{1}{l}{{42.4} (60.8)} &  & {2} (64)      & {46.4} (104.4)                     & {8} (94.4) \\
S27 & {13.2} (108)   & {35.6} (86)                           & \multicolumn{1}{l}{{0} (50.4)}  &                     & {-7.6} (55.2) & {-9.6} (70.4)                      & {0} (54.8)                  \\   
\bottomrule
\end{tabular}}
\label{tab:realdatatde}
\end{table*}

\section{Discussion} \label{sec:discussion}

\noindent In this paper, we propose two novel TDE approaches based on cross- and bispectral analysis, which are insensitive to linearly and instantaneously mixed noise sources. We further provide a versatile testing framework within which we show that our approaches consistently outperform conventional TDE approaches in mixed-noise settings. We also demonstrate that TDE in electrophysiological data is in feasible even in absence of known timings of external stimuli.


\subsection{Robustness to Mixed Noise artifacts}

\noindent Current cross-spectrum and bispectrum based TDE approaches are generally not reliable in presence of mixed noise. For cross-spectrum based TDE, where a delay estimate is derived from the slope of the phase spectrum, mixed noise biases this estimate by flattening the phase spectrum. In bispectral TDE, noise mixing produces spurious zero-delay peaks with possible sidelobes in the bispectral hologram. We show that, using the periodicity of the phase spectrum rather than its slope, and using the sign of the PSI as a measure of directionality between two channels, we can mitigate mixed noise biases in cross-spectrum based TDE. Similarly, we show that spurious zero-delay estimates in bispectral TDE can be successfully removed by antisymmetrization of the bispectrum. Both proposed methods remove biases introduced by mixed noise and allow for TDE in unprecedentedly low SNR regimes. Interestingly, while the conventional non-robust metrics exhibit substantial, expected, TDE performance drops in the mixed noise settings, performance markedly increased in the mixed noise compared to the unmixed setting for our novel robust estimators.
The performance of our methods remained unaltered when introducing auto-correlation to the noise (see Supplementary Section~\ref{sec:expcolor}). Mixing of auto-correlated time-series can introduce further, nonzero spurious peaks, which renders cross-correlational TDE infeasible under these conditions. However, our results demonstrate that, in the presence of (a moderate degree of) noise auto-correlation, antisymmetrized bispectrum based and cross-spectral phase-periodicity based TDE approaches still produce reliable time delay estimates. 

\subsection{Time-Delay Estimation with Gaussian Sources}

\noindent We further report that our implementation of bispectrum based TDE remains sensitive to contributions from Gaussian sources. This is surprising insofar one previously demonstrated theoretical property of bispectral TDE is its robustness to Gaussian sources and thus its ability to reliably estimate delays in mixed Gaussian noise regimes \cite{Nikias1988}. Instead, we found our bispectral TDE approaches to be sensitive to mixed Gaussian noise interactions.

\cite{Nikias1988} already pointed out differences in noise suppression of different numerical approximations of the bispectrum. Specifically, they observe Gaussian suppression to be more effective if the bispectrum is calculated indirectly by taking the Fourier transform of the third-order moments rather than by computing the bispectrum directly from a Fourier transformed time series (see Eq.~\eqref{eq:bispectrum}). We illustrate the different outputs of direct and indirect bispectrum approximations in Supplement~\ref{supp:oldvsnewbispec}). In both cases, the bispectrum theoretically converges to zero for Gaussian sources. However, whereas \cite{Nikias1988} report reduced Gaussian suppression in the direct FFT approximation of the bispectrum, we do not observe suppression of Gaussian sources at all. While the mechanisms behind this effect remain to be clarified, we presume that, while the absolute value of the bispectrum converges the zero, the phase information used for TDE retains information about the underlying instantaneous delay of the noise. This matter should be subject to future investigation.

Yet, we argue that the direct FFT approach to computing the bispectrum provides an advantage over the indirect approach in the context of TDE. This way, it is possible to perform TDE even for Gaussian signals and Gaussian noise combinations and it is thus not required to make any assumptions about the statistical properties of the underlying sources. Contributions from mixed Gaussian noise may not be filtered out this way, but can be filtered out subsequently by means of antisymmetrization. We show that antisymmetrization does not substantially compromise the sensitivity or accuracy of TDE, and again can be applied to mixed noise environments, regardless of the properties of the underlying sources. 

\subsection{Two-Step TDE approach Against Uncertainties}

\noindent We further show that it is possible to effectively measure delay estimate uncertainties using a bootstrapping procedure. All TDE approaches discussed in this paper are forced-choice by nature, which makes it important to discern reliable from unreliable delay estimates. In Section~\ref{sec:expGauss}, we show that conventional TDE approaches produce spurious zero-delay estimates when presented with non-interacting time series or in the presence of mixed noise, with no means of identifying these biases without any knowledge of the underlying sources. Our proposed TDE approaches effectively remove these false positive reports of zero delays. We combine these properties of conventional and novel TDE approaches into a statistical test which successfully filters out false results. Combining this filtering step with actual delay quantification, our proposed TDE approaches incur TDE errors that are consistently close to zero. Within our framework, we find all cross- and bispectrum based two-step TDE approaches to be reliable, although we observe a clear advantage of the novel proposed methods in terms of sensitivity in mixed-noise settings.

\subsection{Unified Testing Framework}
\noindent Apart from presenting two novel TDE approaches, we provide a well defined framework in which we compare their performance to that of conventional TDE methods. As previous works pointed out an advantage of bispectrum based TDE over cross-spectrum based methods in Gaussian mixed noise environments specifically, we aimed to resolve their differences with respect to this property. So far, no studies exist that compare cross- and bispectrum based TDE approaches. With the present work, we provide a comprehensive overview of the behaviour of each method (i) under mixed and unmixed noise influence, (ii) for Gaussian and non-Gaussian distributed noise, (iii) for noise sources with a controlled degree of auto-correlation, and (iv) under different SNRs. In terms of accuracy, we show that both phase-periodicity and antisymmetrized bispectrum based TDE perform equally well throughout our testing environments and consistently outperform their conventional counterparts under mixed noise. 

\subsection{Viability of TDE Analysis on EEG Data}
\noindent We were able to quantify time delays between peripheral median nerve and cortical electrode recordings in two out of three subjects with our ASB TDE approach and one out of three subjects with a phase-periodicity based TDE approach. For the subject where no TDE outputs could be obtained, we hypothesize a too low signal amplitude to be the reason. Overall, the values obtained through our TDE approaches are comparable to the latencies recorded between the peripheral electrode response peak and the evoked cortical N20 peak at $\approx 13$ ms. We see the closeness of results between these approaches as a strong argument for the biological plausibility of our TDE outputs. 

The delays produced by the ASB and phase periodicity approaches did not fully coincide, with marginal differences ranging from 1.8 to 2.4 ms for the same electrode pairings between approaches, corresponding to two and three time bins, respectively. We presume that ASB and phase periodicity based TDE inherently capture different features of the signal and, when confronted with complex data such as biological signals, may produce different results. 

Additionally, we found similar results between TDE analyses conducted on trials time-locked to stimulus onset and trials  obtained by arbitrary segmentation of the continuous data. Whenever TDE outputs of both segmentation approaches passed our confidence filter, their values were identical. This is remarkable insofar that, with TDE analysis, we are presenting a method that is able to extract delay information between channels without any prior information about the underlying signal. This feature represents a significant advantage over traditional ERP-based signal latency extraction, notably in paradigms where latency, rather than response shape, is of primary interest. Moreover, these methods offer a scalable solution for analyzing large datasets without the need for manual intervention, making them highly suitable for high-throughput studies.

\subsection{Limitations}

\noindent The present study also comes with limitations. While our novel cross- and bispectrum based TDE approaches show desirable performance in our simulated environments and in some of the assessed EEG trials, their general applicability will need to be confirmed with more data. It is of particular interest how additional complexities such interactions limited to specific frequency band affect TDE outcomes. In the case of band-limited TDE analysis, for example, it will be necessary to determine a minimum bandwidth for which TDE is still expected to work. For the phase-periodicity estimate, this is determined by minimum bandwidth that still captures at least one period of the sawtooth shaped phase spectrum. Which, in turn, is proportional to the length of the period and inversely proportional to the underlying delay:
\begin{align}
    l_{\text{period}} = \frac{l_{\text{full-band}}}{\tau} \;.
\end{align}
\noindent A bandwidth greater than $ 2 l_{\text{period}}$ needs to be available accordingly in order to resolve a desired minimum delay $\tau$. In contrast, such a hard limit on bandwidth does not exist for bispectrum based TDE. Here, delay information is obtained by integrating across the whole frequency spectrum, potentially with frequency specific weights applied to remove the influence of `silent' regions of the bispectral plane (c.f. Eqs.~\eqref{eq:hologram2},~\eqref{eq:hologram4}). Still, it can be expected that narrowing the band of interest will decrease SNR, until TDE eventually breaks down. How this potential limitation of bispectrum based TDE compares to the minimum bandwidth requirement in phase-periodicity based TDE is yet to be studied. Future work will determine whether environments exist in which one approach should be preferred over the other.

\subsection{Outlook}

\noindent The robustness of our novel TDE methods to mixing artifacts make them promising candidates for the analysis of real-world data. Here we specifically consider the case of brain electrophysiological data, where source mixing and common referencing are known to be the major obstacles to brain functional connectivity analysis \cite{Nolte2005,Nolte2008,Haufe2013,Bastos2016}. Extended testing of TDE approaches on (mock) physiological signals, such as spike trains or EEG time series could further support the usage of TDE as a robust metric to describe functional relationships. In particular, we envision to benchmark TDE efficacy for well-defined brain sources using realistic EEG forward modeling techniques.

In the present study, we assumed a delay between two measurement channels. However, in practical applications such as EEG recordings, the delayed interaction often occurs at the level of unobservable sources, while the recorded channels will comprise linear combinations of these sources. 
For this setting,  we identify two main questions for future research: (i) How viable is TDE for the characterization of source level functional relationships from reconstructed sources activity, and (ii) how can the activity of multiple sources possibly interacting through distinct time delays be accounted for in TDE approaches. The latter in particular poses a considerable limitation to existing TDE methods, as they all rely on finding a single maximal value within a spectrum. It will be necessary to extend TDE quantification methods to account for multiple interactions. Additionally, we stress that our confidence filter of delay estimates also assumes only one underlying delay. We see here the potential for development of new approaches for evaluating the phase spectrum and the bispectral hologram.

\section{Conclusion}

Time delay estimation with conventional methods can lead to spurious results in mixe dnoise settings. To address this, we propose two novel TDE approaches robust to such artifacts, one based on the cross-spectrum, which estimates the periodicity of the phase spectrum and uses additional information about the directionality of information flow via the Phase Slope Index, and another one based on the bispectral hologram, which utilizes antisymmetrization of bispectra to mitigate influences from mixtures of independent sources. Our methods perform best when combined with a bootstrapping procedure to filter out low-confidence estimates and can achieve close to zero-error accuracy. 
These results are independent of the statistical properties of the underlying signal and noise components that were tested, suggesting that our approaches are fit for real-world usage in settings involving source mixing and shared electrical references. We provide a first example of a small electrophysiological test paradigm, where TDE analysis appeared to be effective. Future work will need to rigorously characterise the behaviour of our methods in environments of increasing complexity.



\subsection{Acknowledgments}
\noindent
This result is part of a project that has received funding from the European Research Council (ERC) under the European Union’s Horizon 2020 research and innovation programme (Grant agreement No. 758985). We thank Jan-Mathijs Schoffelen for fruitful discussions.

\bibliographystyle{IEEEtran}
\bibliography{thesis_bib}

\begin{thebibliography}{10}
\providecommand{\url}[1]{#1}
\csname url@samestyle\endcsname
\providecommand{\newblock}{\relax}
\providecommand{\bibinfo}[2]{#2}
\providecommand{\BIBentrySTDinterwordspacing}{\spaceskip=0pt\relax}
\providecommand{\BIBentryALTinterwordstretchfactor}{4}
\providecommand{\BIBentryALTinterwordspacing}{\spaceskip=\fontdimen2\font plus
\BIBentryALTinterwordstretchfactor\fontdimen3\font minus \fontdimen4\font\relax}
\providecommand{\BIBforeignlanguage}[2]{{%
\expandafter\ifx\csname l@#1\endcsname\relax
\typeout{** WARNING: IEEEtran.bst: No hyphenation pattern has been}%
\typeout{** loaded for the language `#1'. Using the pattern for}%
\typeout{** the default language instead.}%
\else
\language=\csname l@#1\endcsname
\fi
#2}}
\providecommand{\BIBdecl}{\relax}
\BIBdecl

\bibitem{Knapp1976}
C.~Knapp and G.~Carter, ``{The generalized correlation method for estimation of time delay},'' \emph{IEEE Transactions on Acoustics, Speech, and Signal Processing}, vol.~24, no.~4, pp. 320--327, aug 1976.

\bibitem{Nikias1988}
C.~L. Nikias and R.~Pan, ``{Time Delay Estimation in Unknown Gaussian Spatially Correlated Noise},'' \emph{IEEE Transactions on Acoustics, Speech, and Signal Processing}, vol.~36, no.~11, pp. 1706--1714, 1988.

\bibitem{Chella2014}
F.~Chella, L.~Marzetti, V.~Pizzella, F.~Zappasodi, and G.~Nolte, ``{Third order spectral analysis robust to mixing artifacts for mapping cross-frequency interactions in EEG/MEG},'' \emph{NeuroImage}, vol.~91, no.~1, pp. 146--161, may 2014.

\bibitem{ZhaoZhen1984}
{Zhao Zhen} and {Hou Zi-qiang}, ``{The generalized phase spectrum method for time delay estimation},'' in \emph{ICASSP '84. IEEE International Conference on Acoustics, Speech, and Signal Processing}, vol.~9, no.~3.\hskip 1em plus 0.5em minus 0.4em\relax Institute of Electrical and Electronics Engineers, 1984, pp. 459--462.

\bibitem{Nikias1993}
C.~L. Nikias, J.~M. Mendel, and A.~Swami, \emph{{Higher-Order Spectral Analysis Matlab Toolbox User's Guide}}, 2nd~ed.\hskip 1em plus 0.5em minus 0.4em\relax United Signals \& Systems, Inc, 1993.

\bibitem{Stam2007}
C.~J. Stam, G.~Nolte, and A.~Daffertshofer, ``{Phase lag index: Assessment of functional connectivity from multi channel EEG and MEG with diminished bias from common sources},'' \emph{Human Brain Mapping}, vol.~28, no.~11, pp. 1178--1193, 2007.

\bibitem{Brennan2007}
M.~J. Brennan, Y.~Gao, and P.~F. Joseph, ``{On the relationship between time and frequency domain methods in time delay estimation for leak detection in water distribution pipes},'' \emph{Journal of Sound and Vibration}, vol. 304, no. 1-2, pp. 213--223, 2007.

\bibitem{Faerman2022}
V.~Faerman, V.~Avramchuk, K.~Voevodin, I.~Sidorov, and E.~Kostyuchenko, ``{Study of Generalized Phase Spectrum Time Delay Estimation Method for Source Positioning in Small Room Acoustic Environment},'' \emph{Sensors}, vol.~22, no.~3, 2022.

\bibitem{Salmelin2009}
R.~Salmelin and S.~Baillet, ``{Electromagnetic brain imaging},'' \emph{Human Brain Mapping}, vol.~30, no.~6, pp. 1753--1757, 2009.

\bibitem{Nolte2004}
G.~Nolte, O.~Bai, L.~Wheaton, Z.~Mari, S.~Vorbach, and M.~Hallett, ``{Identifying true brain interaction from EEG data using the imaginary part of coherency},'' \emph{Clinical Neurophysiology}, vol. 115, no.~10, pp. 2292--2307, 2004.

\bibitem{Nolte2005}
G.~Nolte, A.~Ziehe, F.~Meinecke, and K.~R. M{\"{u}}ller, ``{Analyzing coupled brain sources: Distinguishing true from spurious interaction},'' \emph{Advances in Neural Information Processing Systems}, vol.~c, pp. 1027--1034, 2005.

\bibitem{Nolte2008}
G.~Nolte, A.~Ziehe, V.~V. Nikulin, A.~Schl{\"{o}}gl, N.~Kr{\"{a}}mer, T.~Brismar, and K.~R. M{\"{u}}ller, ``{Robustly estimating the flow direction of information in complex physical systems},'' \emph{Physical Review Letters}, vol. 100, no.~23, pp. 1--4, 2008.

\bibitem{Vinck2011}
M.~Vinck, R.~Oostenveld, M.~{Van Wingerden}, F.~Battaglia, and C.~M. Pennartz, ``{An improved index of phase-synchronization for electrophysiological data in the presence of volume-conduction, noise and sample-size bias},'' \emph{NeuroImage}, vol.~55, no.~4, pp. 1548--1565, 2011.

\bibitem{Bastos2016}
A.~M. Bastos and J.~M. Schoffelen, ``{A tutorial review of functional connectivity analysis methods and their interpretational pitfalls},'' \emph{Frontiers in Systems Neuroscience}, vol.~9, no. JAN2016, pp. 1--23, 2016.

\bibitem{Winkler2016}
I.~Winkler, D.~Panknin, D.~Bartz, K.~R. Muller, and S.~Haufe, ``{Validity of Time Reversal for Testing Granger Causality},'' \emph{IEEE Transactions on Signal Processing}, vol.~64, no.~11, pp. 2746--2760, 2016.

\bibitem{Baillet2020}
S.~Baillet, ``{Encyclopedia of Computational Neuroscience},'' in \emph{Encyclopedia of Computational Neuroscience}, D.~Jaeger and R.~Jung, Eds.\hskip 1em plus 0.5em minus 0.4em\relax New York, NY: Springer New York, 2020, no. January 2014, ch. 529.

\bibitem{Pellegrini2023}
F.~Pellegrini, T.~D. Nguyen, T.~Herrera, V.~Nikulin, G.~Nolte, and S.~Haufe, ``{Distinguishing between- from within-site phase-amplitude coupling using antisymmetrized bispectra},'' \emph{bioRxiv}, 2023.

\bibitem{Piersol1981}
A.~G. Piersol, ``{Time Delay Estimation Using Phase Data},'' \emph{IEEE Transactions on Acoustics, Speech, and Signal Processing}, vol.~29, no.~3, pp. 471--477, 1981.

\bibitem{Buzsaki2011}
G.~Buzs{\'{a}}ki and A.~Draguhn, ``{Neuronal Oscillations in Cortical Networks},'' \emph{Advancement Of Science}, vol. 304, no. 5679, pp. 1926--1929, 2011.

\bibitem{Aru2015}
J.~Aru, J.~Aru, V.~Priesemann, M.~Wibral, L.~Lana, G.~Pipa, W.~Singer, and R.~Vicente, ``{Untangling cross-frequency coupling in neuroscience},'' \emph{Current Opinion in Neurobiology}, vol.~31, pp. 51--61, 2015.

\bibitem{Chen2011}
A.~K. Engel, P.~Fries, and W.~Singer, ``{Dynamic predictions: Oscillations and synchrony in top–down processing},'' \emph{Nature Reviews Neuroscience}, vol.~2, no.~10, pp. 704--716, oct 2001.

\bibitem{Isserlis1918}
L.~Isserlis, ``{On a Formula for the Product-Moment Coefficient of any Order of a Normal Frequency Distribution in any Number of Variables},'' \emph{Biometrika}, vol.~12, no. 1/2, p. 134, 1918.

\bibitem{allison1991potentials}
T.~Allison, G.~McCarthy, C.~C. Wood, and S.~J. Jones, ``Potentials evoked in human and monkey cerebral cortex by stimulation of the median nerve: a review of scalp and intracranial recordings,'' \emph{Brain}, vol. 114, no.~6, pp. 2465--2503, 1991.

\bibitem{cruccu2008recommendations}
G.~Cruccu, M.~Aminoff, G.~Curio, J.~Guerit, R.~Kakigi, F.~Mauguiere, P.~Rossini, R.-D. Treede, and L.~Garcia-Larrea, ``Recommendations for the clinical use of somatosensory-evoked potentials,'' \emph{Clinical neurophysiology}, vol. 119, no.~8, pp. 1705--1719, 2008.

\bibitem{stephani2020temporal}
T.~Stephani, G.~Waterstraat, S.~Haufe, G.~Curio, A.~Villringer, and V.~V. Nikulin, ``Temporal signatures of criticality in human cortical excitability as probed by early somatosensory responses,'' \emph{Journal of Neuroscience}, vol.~40, no.~34, pp. 6572--6583, 2020.

\bibitem{Zhang2014}
J.~Zhang, X.~Yin, L.~Zhao, A.~C. Evans, L.~Song, B.~Xie, H.~Li, C.~Luo, and J.~Wang, ``{Regional alterations in cortical thickness and white matter integrity in amyotrophic lateral sclerosis},'' \emph{Journal of Neurology}, vol. 261, no.~2, pp. 412--421, 2014.

\bibitem{delorme2004eeglab}
A.~Delorme and S.~Makeig, ``Eeglab: an open source toolbox for analysis of single-trial eeg dynamics including independent component analysis,'' \emph{Journal of neuroscience methods}, vol. 134, no.~1, pp. 9--21, 2004.

\bibitem{Haufe2013}
S.~Haufe, V.~V. Nikulin, K.~R. M{\"{u}}ller, and G.~Nolte, ``{A critical assessment of connectivity measures for EEG data: A simulation study},'' \emph{NeuroImage}, vol.~64, no.~1, pp. 120--133, 2013.

\end{thebibliography}

\clearpage

\setcounter{page}{1}

\pagestyle{plain}

\renewcommand\thefigure{S\arabic{figure}}  
\renewcommand\thesection{S-\Roman{section}}  

\setcounter{figure}{0}
\setcounter{section}{0}
\setcounter{subsection}{0}

\begin{figure*}[htbp]
\centering
\includegraphics[width=.7\textwidth]{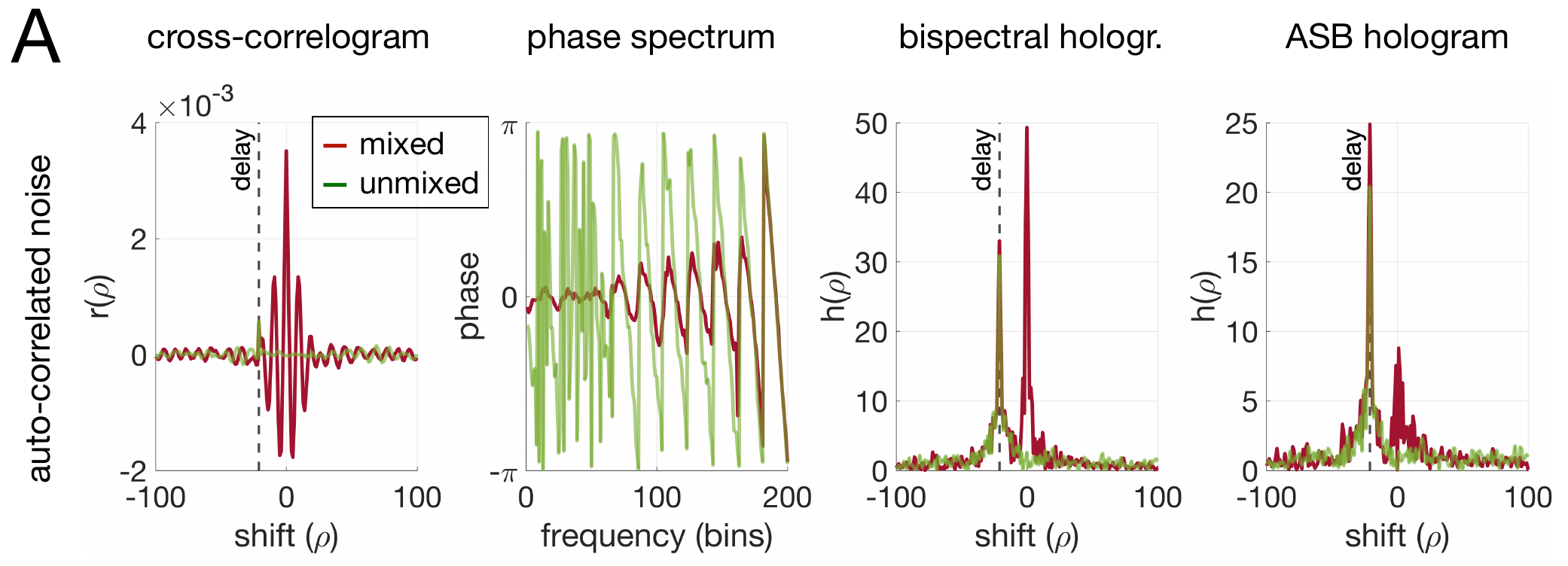}
\includegraphics[width=.7\textwidth]{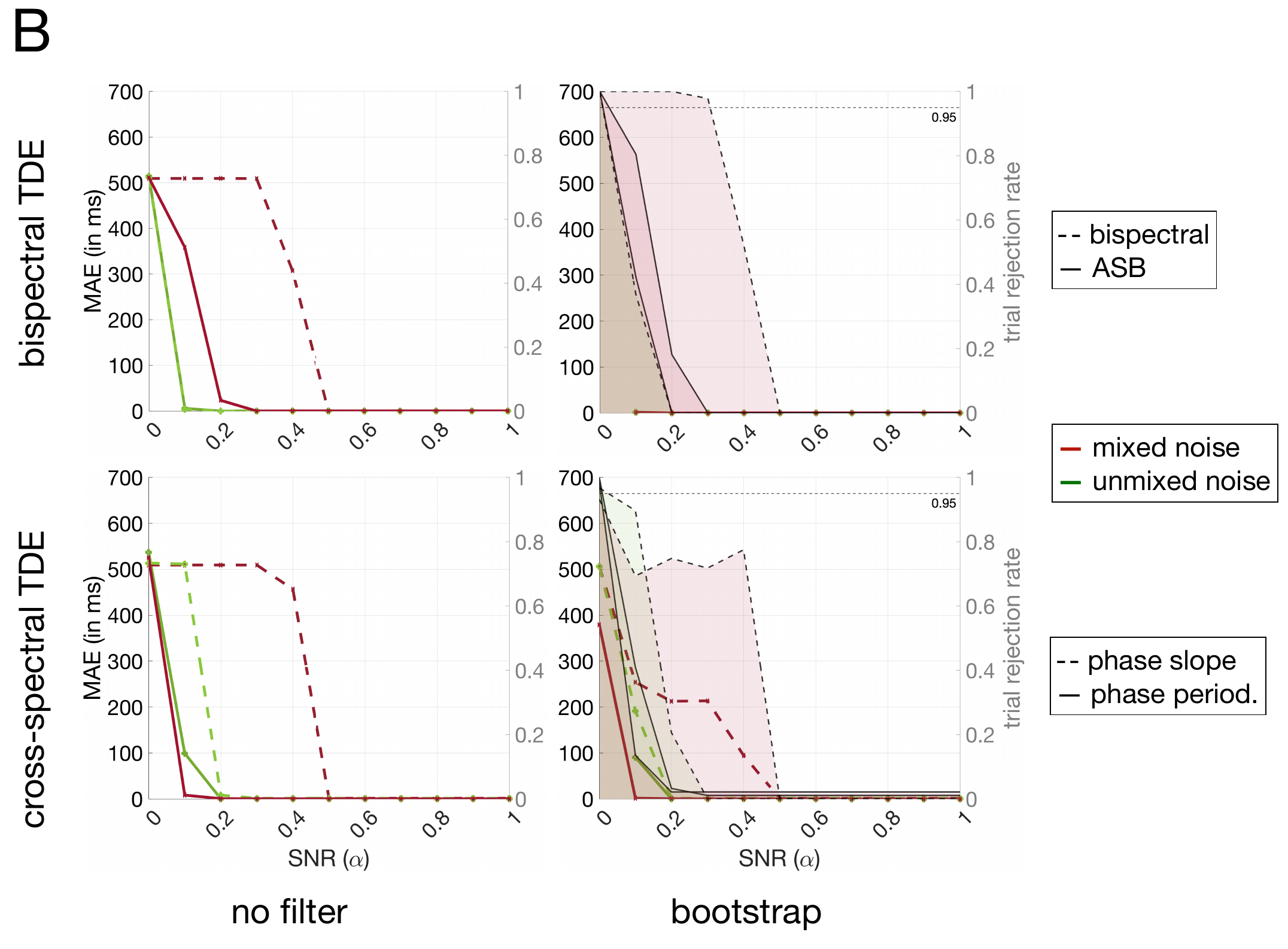}
\includegraphics[width=.7\textwidth]{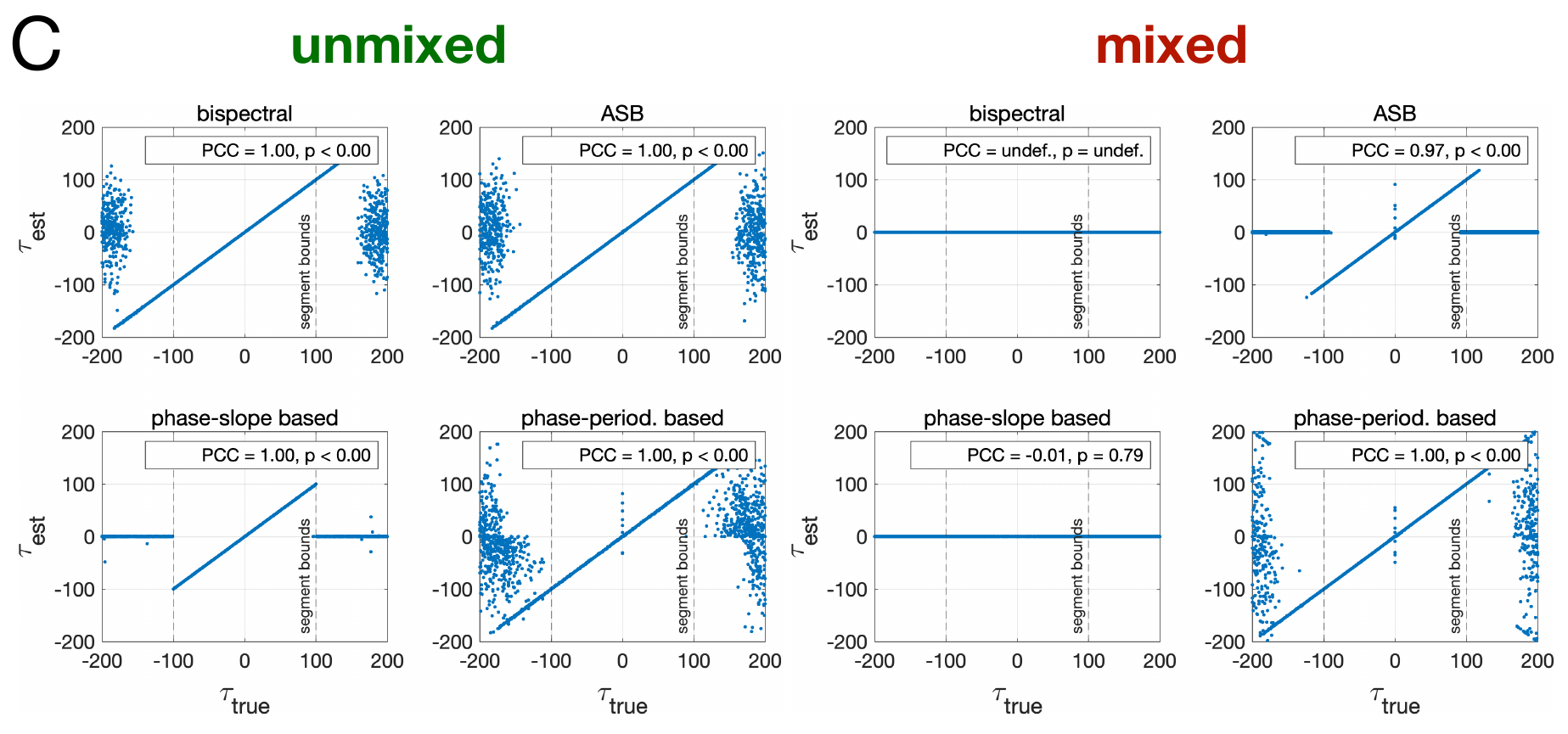}
\caption{Performance of time delay estimation (TDE) approaches in the presence of auto-correlated noise. A: Cross-correlogram, phase spectrum, bispectral hologram, and antisymmetrized bispectral hologramfor exponentially distributed (non-Gaussian) signal and auto-correlated (non-Gaussian) noise at SNR $\alpha = 0.3$. B: Performance of bispectrum and cross-spectrum based TDE at various SNR levels, with (right) and without (left) a bootstrap based filtering approach for low-confidence estimates applied. Shown are the mean absolute errors (MAEs) for $n_{trial} = 500$ trials for each SNR level. C: Distribution of true ($\tau(\text{true})$) versus estimated ($\tau(\text{est})$) delays (n=10 trials per $\tau(\text{true)} = [-200,200]$) for SNR $\alpha = 0.2$. Pearson's correlation coefficient (PCC) and corresponding p-values are reported for true lags contained within the segment bounds $[-100;100]$.}
\label{fig:perfpink}
\end{figure*}

\begin{strip}

\section*{\LARGE \textup{Estimating Time Delays between Signals under Mixed Noise  
Influence with Novel Cross- and Bispectral Methods (Supplementary Material)}}

\begin{center}
Tin Jurhar, Franziska Pellegrini, Ana I. Nuñes del Toro, Tilman Stephani, Guido Nolte, Stefan Haufe
\end{center}

\section{Simulations with Auto-correlated Noise} \label{sec:expcolor}

\noindent In the main text, we established that noise Gaussianity does not substantially affect TD estimation for any of the tested methods, contrary to our expectation for BS based TDE. 

Another variable that can influence TDE is that of noise auto-correlation. So far, we have assumed temporally white processes for both the signal and noise time series, leading to sharp peaks in the auto-correlation spectra and bispectral holograms, essentially representing Dirac delta functions. In most real-world applications, however, measured time series possess inherent temporal dependencies, which are reflected in non-flat power spectra and non-zero auto-correlations that extend over considerable numbers of lags. Especially for periodic activity, it is common to observe that the auto-correlation spectrum has several secondary peaks (side lobes) besides the main peak at lag zero, oscillating in positive and negative direction over a possible long range of lags. The presence of such side lobes can thus complicate TDE in mixed noise settings. For example, such a case would result in not only one, but multiple spurious peaks in the observed cross-correlogram, not all of which will be located at lag zero. Correlational TDE may therefore not be applicable in contexts where (noise) auto-correlation is to be expected. To see how noise auto-correlation affects TDE approaches, we next test their performance in an auto-correlated setting. 

\subsection{Methods}
\noindent We here repeat the experiment outlined in \textit{Section~\ref{sec:expGauss}}, this time considering non-Gaussian auto-correlated noise. We model the auto-correlated noise with two components, a non-periodic pink noise process and a periodic alpha oscillatory component. The pink noise process is generated with power spectral density $PSD(f) = \nicefrac{1}{|f|^{\lambda}}$ imposed, where we choose $\lambda = 0.7$ and randomly sample the phase at each frequency independently from the full interval $[-\pi, \pi]$. The alpha wave component is constructed by sampling Gaussian white noise from a standard normal ($\mathcal{N}(0,1)$), and applying (i) a 1 Hz highpass-, (ii) a 45 Hz lowpass-, and (iii) a $[8, 13]$ Hz Butterworth filter. The periodic and non-periodic components are then power-normalized and added together. We again combine this modeled noise with a signal component identical to the procedure in the previous experiment (Section \ref{sec:expGauss}) and generate our data sets from which we derive MAE metrics. 

\subsection{Results}
\subsubsection{Effects of Noise Auto-Correlation on Cross-Correlogram, Phase Spectrum, and Bispectral Holograms}
\noindent The cross-correlogram, phase spectrum and bispectral holograms for auto-correlated noise combinations at SNR $\alpha = 0.3$ are shown in Figure \ref{fig:perfpink}. For unmixed-noise environments (green) the cross-correlogram is generally flat, with a peak at the value of the underlying delay. For mixed-noise environments (red), the cross-correlogram exhibits its largest maximum at lag zero, with several side lobes to its left and right. In this example, the peak accredited to the underlying delay coincides with one of the side lobes, at $\tau = -21$, rendering correlational TDE infeasible.

For mixed and unmixed noise settings, the phase difference spectra has characteristic sawtooth features. In the unmixed-noise setting and for the lower frequency range, these features appear distorted but overall range within $[-\pi, \pi]$. In the mixed-noise setting, the amplitude of the phase spectrum is indirectly proportional to the underlying frequency, indicative of the $\nicefrac{1}{|f|}$ power spectrum of the modeled noise process. Notably, the periodicity of phase spectrum remains conserved.

For the bispectral hologram, we again observe an additional peak at lag zero for the  mixed-noise setting. Bispectral antisymmetrization again appears to be largely successful at suppressing this contribution; however, a portion of the initial signal remains. Results described next show to what extent this influences TDE efficacy in presence of auto-correlated noise.

\subsubsection{Accuracy of TDE approaches in Presence of Auto-Correlated Noise}
\noindent Mean absolute errors for auto-correlated noise with no bootstrapping filter applied (left panels of Figure~\ref{fig:perfpink}B) are comparable to those of the previous experiment. In unmixed noise settings (green), the MAE of conventional and antisymmetrized TDE estimates are comparable, with MAE $\approx 500$ ms for SNR $\alpha = 0$ and MAE $\approx$ 0 for SNR $\alpha \geq 0.1$. For cross-spectrum based TDE, our phase-periodicity approach performed slightly better than its phase-slope based counterpart. In mixed-noise settings (red), the MAEs of conventional TDE approaches remained steadily high at $\approx 500$ ms, until SNR $\alpha = 0.3$ and, then flattening to $\approx 0$ at SNR $\alpha \geq 0.5$. The MAEs of our ASB and phase-periodicity TDE approach are comparable between mixed noise and unmixed noise settings, remaining near zero for all SNR $\alpha \geq 0.2$.

When applying the proposed confidence filter to the TDE estimates, we observe that, across all trials, the acceptance rate is below the 5\% threshold for SNR $\alpha = 0$ (right panels of Figure \ref{fig:perfpink}B). For both conventional and antisymmetrized bispectral TDE, our filter reliably rejects all trials with false TDE outputs. This is reflected by a trial rejection rate close to 1 for low SNR regimes, and, moreover, an MAE of zero of all accepted trials. For our cross-spectrum based TDE approaches, filtering did not remove all false outputs. In the unmixed noise setting, the MAEs of phase-slope and phase-periodicity are $\approx 40$ ms and $\approx 100$ ms, respectively, with trial rejection rates of 0.92 and and 0.14 at SNR $\alpha = 0.1$. At higher SNR, this error reduces to zero. For mixed noise trials, we report a nonzero MAE for our phase-periodicity approach at SNR $\alpha = 0$. However, this value is accredited to one single falsely accepted trial. Otherwise phase-periodicity based TDE in combination with a filtering approach produced near-zero MAEs. Compared to this, MAEs for phase-slope TD estimates are comparatively high. For low SNR regimes, the MAE is around 300 ms, and eventually drops to zero for SNR $\alpha \geq 0.5$. The behaviour of TDE approaches between environments without and with noise auto-correlation in terms of MAE is thus largely comparable. In environments with (controlled) noise auto-correlation, TDE thus still appears feasible.

\subsubsection{Distribution of TD Estimates for Signals with Noise Auto-Correlation}

\noindent Figure~\ref{fig:perfpink}C summarizes the distributions of delay estimates for cross- and bispectrum TDE approaches for SNR $\alpha = 0$ and $\alpha = 0.2$. The distributions of TDE outputs from trials with noise auto-correlation are comparable to the Gaussian and exponential noise cases, with one notable difference: When no signal is present at all, (SNR $\alpha = 0$) our phase-periodicity TDE approach estimates more delays to lie at the interval bounds $[-200,200]$, in both unmixed and mixed noise settings. This is discernible from a thin band of data points at the top and bottom of each phase-periodicity distribution plot. The effect is observed throughout the entire range of possible values, and does not seem to depend of the magnitude of the underlying delay. 

\section{Derivations}

\subsection{Mixed and Unmixed Noise Contributions to The Cross-Correlogram}
Here, we derive the cross-correlogram for the simple model Eq.~\eqref{eq:corr}. After expansion, we can disentangle signal and noise contributions to the observed cross-correlogram. For mixed noise environments $(\theta_{1,2}\neq 0$), we obtain:
\begin{align}
\nonumber r_{\text{mixed}}(\rho) =& \; E\left[ X(t) Y(t+\rho) \right] \\
\nonumber=& \; E\left[  \left( \alpha x(t)+ (1- \alpha)\left[n_X(t)+\theta_1 n_Y(t)\right]\right) \right. \ldots\\
\nonumber&\left.\left(\alpha \beta x(t-\tau+\rho)+(1-\alpha)\left[ n_Y(t+\rho)+\theta_2 n_X(t+\rho)\right]\right) \right] \\
\nonumber =& \; E\left[ \alpha^2 \beta x(t)  x(t-\tau+\rho) + (\alpha-\alpha^2)x(t)(n_Y(t+\rho)+\theta_2n_X(t+\rho))\right.\ldots \\
\nonumber &+ (\alpha-\alpha^2)\beta x(t-\tau+\rho)(n_X(t)+\theta_1n_Y(t)) \ldots \\
\nonumber &+ \left.(1-\alpha)^2 \big(n_X(t)n_Y(t+\rho)+\theta_2n_X(t)n_X(t+\rho)+\theta_1n_Y(t)n_Y(t+\rho)+\theta_1\theta_2 n_X(t)n_Y(t+\rho)\big)\right] \\
\nonumber =& \; \alpha^2\beta E\left[ x(t) x(t-\tau+\rho)\right] + (\alpha-\alpha^2)\left(E\left[x(t)n_Y(t+\rho)\right]+\theta_2E\left[x(t)n_X(t+\rho)\right]\right)\ldots \\
\nonumber &+ (\alpha-\alpha^2)\beta\left(E\left[ x(t-\tau+\rho)n_X(t)\right]+\beta\theta_1E\left[x(t-\tau+\rho)n_Y(t)\right]\right) \ldots \\
\nonumber &+(1-\alpha)^2 \big(E\left[n_X(t)n_Y(t+\rho)\right]+\theta_2E\left[n_X(t)n_X(t+\rho)\right]\ldots\\
\nonumber &+\theta_1E\left[n_Y(t)n_Y(t+\rho)\right]+\theta_1\theta_2E\left[n_X(t)n_Y(t+\rho)\big)\right] \\
 =& \; \alpha^2\beta E\left[ x(t) x(t-\tau+\rho)\right]+\theta_2E\left[n_X(t)n_X(t+\rho)\right]+\theta_1E\left[n_Y(t)n_Y(t+\rho)\right] \;.
 \label{supp:corrmixed}
 \end{align}

The resulting cross-correlogram for mixed noise settings is thus a superposition of a signal and two noise cross-correlograms. By setting $\theta_{1,2} = 0$ in Eq.~\eqref{supp:corrmixed}, we find that, for unmixed noise time-series, the cross-correlogram reduces to:
\begin{align}
 r_{\text{unmixed}}(\rho) = & \alpha^2\beta E\left[ x(t) x(t-\tau+\rho)\right] \label{supp:corrunmixed} \;.
\end{align}

\subsection{Mixed and Unmixed Noise Contributions to The Phase Spectrum}

Similarly, we can expand and decompose Eq.~\eqref{eq:phasespec} into the respective signal and noise Fourier coefficients of $X(t),Y(t)$ to investigate the effect of noise mixing on the phase spectrum:
%
\begin{align} 
\nonumber     P_{XY\text{,mixed}} =& \; \angle(S_{XY} = \angle(\langle F_XF_Y^*\rangle)\\
\nonumber = & \; \angle(\langle \left(\alpha F_x + (1-\alpha)(F_{n_X}+\theta_1F_{n_Y})\right)\left(\alpha \beta F_y^*+(1-\alpha)(F_{n_Y}^*+\theta_2F^*_{n_X})\right)\rangle)\\
\nonumber =& \; \angle(\langle\ldots\\
\nonumber     &\alpha^2\beta F_xF_y^*+(\alpha-\alpha^2)(F_xF_{n_Y}^*+\theta_2F_xF_{n_X}^*)+ (\alpha-\alpha^2)\beta (F_y^*F_{n_X}+\theta_1F^*_yF_{n_Y}) \ldots\\
\nonumber     &+(1-\alpha)^2\left(F_{n_X}F_{n_Y}^*+\theta_2F_{n_X}F_{n_X}*+\theta_1F_{n_Y}F_{n_Y}^*+\theta_1\theta_2F_{n_Y}F_{n_X} \right) \ldots\\
\nonumber     &\ldots\rangle)\\
\nonumber =& \; \angle(\langle\ldots\\
\nonumber     &\alpha^2\beta F_xF_y^*+(\alpha-\alpha^2)(F_xF_{n_Y}^*+\theta_2F_xF_{n_X}^*)+ (\alpha-\alpha^2)\beta (F_y^*F_{n_X}+\theta_1F^*_yF_{n_Y}) \ldots\\
\nonumber     &+(1-\alpha)^2\left(F_{n_X}F_{n_Y}^*+\theta_2F_{n_X}F_{n_X}*+\theta_1F_{n_Y}F_{n_Y}^*+\theta_1\theta_2F_{n_Y}F_{n_X} \right) \ldots\\
\nonumber     &\ldots\rangle)\\
\nonumber =& \; \angle(\langle\alpha^2\beta F_xF_y^*\rangle+(\alpha-\alpha^2)(\langle F_xF_{n_Y}^*\rangle+\theta_2\langle F_xF_{n_X}^*\rangle)+ (\alpha-\alpha^2)\beta (\langle F_y^*F_{n_X}\rangle+\theta_1\langle F^*_yF_{n_Y}\rangle) \ldots\\
\nonumber     &+(1-\alpha)^2(\langle F_{n_X}F_{n_Y}^*\rangle+\theta_2\langle F_{n_X}F_{n_X}^*\rangle+\theta_1\langle F_{n_Y}F_{n_Y}^*\rangle+\theta_1\theta_2\langle F_{n_Y}F_{n_X}\rangle))\\
\nonumber =& \; \angle(\langle\alpha^2\beta F_xF_y^*\rangle + (1-\alpha)^2(\theta_2\langle F_{n_X}F_{n_X}^*\rangle+\theta_1\langle F_{n_Y}F_{n_Y}^*\rangle) \\
=& \angle(\alpha^2\beta \langle F_xF_y^*\rangle + (1-\alpha)^2(\theta_2\langle|F_{n_X}|^2\rangle+\theta_1\langle|F_{n_Y}|^2\rangle) \;.\label{supp:phasemixed}
\end{align}

Here, the dependency on frequency of $P_{XY}$, $S_{XY}$, and Fourier coefficients $F$ is implicit. For the unmixed noise case ($\theta_{1,2} = 0$), the noise contributions vanish:
\begin{align}
P_{XY\text{,unmixed}} =& \angle(\alpha^2\beta\langle F_xF_y^*\rangle) \;.
\label{supp:phaseunmixed}
\end{align}

\subsection{Phase Information In $I_{M2}$,$I_{M4}$ Bispectrum Based TDE Approaches}

Similar to our workings in Eq.~\eqref{eq:phasetodelay3}, we wish to show that the phase information contained in the bispectral hologram for methods $I_{M2}$, $I_{M4}$ corresponds to a combination of phase spectra evaluated for the three frequency components $f_1$,$f_2$,$f_1+f_2$:
\begin{align} 
\nonumber     &\angle(I_{M2}) = \angle(I_{M4}) = \angle(B_{XYX}(f_1,f_2))-\frac{1}{2} \left(\angle(B_{XXX}(f_1,f_2)+\angle(B_{YYY}(f_1,f_2))\right) \\
\nonumber     &=\langle \varphi_X(f_1)+\varphi_Y(f_2)-\varphi_X(f_1+f_2)\rangle - \frac{1}{2}\left(\langle \varphi_X(f_1)+\varphi_X(f_2)-\varphi_X(f_1+f_2)\rangle + \langle \varphi_Y(f_1)+\varphi_Y(f_2)-\varphi_Y(f_1+f_2)\rangle\right) \\
\nonumber     &=\langle \varphi_X(f_1)+\varphi_Y(f_2)-\varphi_X(f_1+f_2)\rangle - \frac{1}{2}\langle \varphi_X(f_1)+\varphi_X(f_2)-\varphi_X(f_1+f_2)\rangle - \frac{1}{2} \langle \varphi_Y(f_1)+\varphi_Y(f_2)-\varphi_Y(f_1+f_2)\rangle \\
\nonumber     &=\langle\frac{1}{2}\varphi_X(f_1)+\frac{1}{2}\varphi_Y(f_2)-\frac{1}{2}\varphi_X(f_1+f_2)-\frac{1}{2}\varphi_X(f_2)-\frac{1}{2}\varphi_Y(f_1)-\frac{1}{2}\varphi_Y(f_1+f_2)\rangle\\
    &= \frac{1}{2} \langle\varphi_{X}(f_1)-\varphi_{Y}(f_1)\rangle - \frac{1}{2} \langle\varphi_{X}(f_2)-\varphi_{Y}(f_2)\rangle - \frac{1}{2} \langle\varphi_{X}(f_1+f_2)-\varphi_{Y}(f_1+f_2)\rangle \;. \label{eq:IM2,IM4phase2}
\end{align} 
The phase information of $I_{M2}$, $I_{M4}$  can thus be viewed as a superposition of phase spectra of the individual frequency components.

\subsection{Composition of The Bispectrum for Additive and Independent Signal and Noise} \label{sec:additivebispectrum}

\noindent Here we show that the bispectrum of timeseries composed of independent additive signal and noise components can be represented as the superposition of two separate signal and noise bispectra. Given timeseries $A,B,C$ with zero-mean signal and noise components, and their respective Fourier coefficients $F_A=F_a+F_{n_A}$ (and similarly for $F_B,F_C$), the resulting bispectrum (Eq.~\eqref{eq:bispectrum}) is 
%
\begin{align} 
\nonumber     B_{ABC}(f_1,f_2) =& \; \left\langle F_A(f_1)F_B(f_2)F_C^*(f_1+f_2)\right\rangle\\
\nonumber     =& \; \left\langle (F_a(f_1)+F_{n_A}(f_1))(F_b(f_2)+F_{n_B}(f_2))(F_c^*(f_1+f_2)+F^*_{n_C}(f_2))\right\rangle\\
 = & \; \langle F_a(f_1)F_b(f_2)F_c^*(f_1+f_2)\rangle + \langle F_a(f_1)F_{n_B}(f_2)F_c^*(f_1+f_2)\rangle +  \langle F_{n_A}(f_1)F_b(f_2)F_c^*(f_1+f_2)\rangle \ldots \label{eq:addbispec1}\\
\nonumber     &+ \langle F_{n_A}(f_1)F_{n_B}(f_2)F_c^*(f_1+f_2)\rangle +\langle F_a(f_1)F_b(f_2)F_{n_C}^*(f_1+f_2)\rangle + \langle F_a(f_1)F_{n_B}(f_2)F_{n_C}^*(f_1+f_2)\rangle \ldots \\
= &+  \langle F_{n_A}(f_1)F_b(f_2)F_{n_C}^*(f_1+f_2)\rangle + \langle F_{n_A}(f_1)F_{n_B}(f_2)F_{n_C}^*(f_1+f_2)\rangle \label{eq:addbispec2}\\
=& \; \langle F_a(f_1)F_b(f_2)F_c^*(f_1+f_2)\rangle + \langle F_{n_A}(f_1)F_{n_B}(f_2)F_{n_C}^*(f_1+f_2)\rangle 
\label{eq:additivebispectrum}
\end{align}
%
\noindent since all terms in \eqref{eq:addbispec1}/\eqref{eq:addbispec2} containing both signal and noise contributions equate to zero when taking the expectation. Thus, similarly to the phase spectrum, noise effects in the bispectrum are also additive. To explore the effect of noise mixing, the noise term in \eqref{eq:additivebispectrum} can be expanded accordingly. We address this case in the following for one timeseries pair and in the context of antisymmetrization only.

\subsection{Mixed Noise Effects on The Antisymmetrized Bispectrum}

\noindent We consider timeseries $X,Y$ and the model Eq.~\eqref{eq:fullTDE}. The respective Fourier coefficients are $F_X = \alpha F_x + (1-\alpha)(F_{n_X} + \theta_1 F_{n_Y})$ and $F_Y = \alpha\beta F_y + (1-\alpha)(F_{n_Y} + \theta_2 F_{n_X})$. In Supplementary Section~\ref{sec:additivebispectrum}, we show that additive noise contributions are also additive in the bispectrum domain. We use this  to show that antisymmetrization removes noise contributions altogether. First, we separate signal and noise terms to assess their individual contributions to the antisymmetrized bispectrum:
%
\begin{align} 
\nonumber     B_{[X|YX]}&(f_1,f_2) = B_{XYX}(f_1,f_2)-B_{YXX}(f_1,f_2) \\
\nonumber     =& \; \langle F_X(f_1)F_Y(f_2)F_X^*(f_1+f_2)\rangle - \langle F_Y(f_1)F_X(f_2)F_X^*(f_1+f_2)\rangle \\
\nonumber     =& \; \langle \left[\alpha F_x(f_1)+(1-\alpha)(F_{n_X}(f_1)+\theta_1F_{n_Y}(f_1))\right] \left[\alpha\beta F_y(f_2)+(1-\alpha)(F_{n_Y}(f_2)+\theta_2F_{n_X}(f_2))\right] \ldots\\
\nonumber     &\cdot\left[\alpha F_x^*(f_1+f_2)+(1-\alpha)(F_{n_X}^*(f_1+f_2)+\theta_1 F_{n_Y}^*(f_1+f_2))\right]\rangle - \langle\left[\alpha\beta F_y(f_1)+(1-\alpha)(F_{n_Y}(f_1)+\theta_2 F_{n_X}(f_1))\right] \ldots \\
\nonumber     &\cdot \left[\alpha F_x(f_2)+(1-\alpha)(F_{n_X}(f_2)+\theta_1 F_{n_Y}(f_2))\right]\left[\alpha\beta F_y(f_2)+(1-\alpha)(F_{n_Y}(f_2)+\theta_2 F_{n_X}(f_2))\right]\rangle\\
\nonumber     =& \; \alpha^3\beta\langle F_x(f_1)F_y(f_2)F_x^*(f_1+f_2)\rangle  \ldots \\ 
\nonumber &+ (1-\alpha)^3\langle(F_{n_X}(f_1)+\theta_1 F_{n_Y}(f_1))(F_{n_Y}(f_2)+\theta_2 F_{n_X}(f_2))(F_{n_X}^*(f_1+f_2)+\theta_1 F_{n_Y}^*(f_1+f_2))\rangle + c.t. \ldots \\
\nonumber     &- \alpha^3\beta\langle F_y(f_1)F_x(f_2)F_x^*(f_1+f_2)\rangle \ldots \\
    &- (1-\alpha)^3\langle(F_{n_Y}(f_1)+\theta_2 F_{n_X}(f_1))(F_{n_X}(f_2)+\theta_1 F_{n_Y}(f_2))(F_{n_X}^*(f_1+f_2)+\theta_1 F_{n_Y}^*(f_1+f_2))\rangle + c.t. \;.
    \label{eq:antisym2}
\end{align}

\noindent Here, $c.t.$ denote terms containing both signal and noise contributions, which evaluate to zero. Eq.~\eqref{eq:antisym2} thus also contains summations of bispectra arising from either signal or noise sources. In the next step, we expand the two noise terms to show that their difference is equal to zero:
%
\begin{align}
\nonumber (1&-\alpha)^3\langle(F_{n_X}(f_1)+\theta_1 F_{n_Y}(f_1))(F_{n_Y}(f_2)+\theta_2 F_{n_X}(f_2))(F_{n_X}^*(f_1+f_2)+\theta_1 F_{n_Y}^*(f_1+f_2))\rangle \ldots\\
\nonumber &-(1-\alpha)^3\langle(F_{n_Y}(f_1)+\theta_2 F_{n_X}(f_1))(F_{n_X}(f_2)+\theta_1 F_{n_Y}(f_2))(F_{n_X}^*(f_1+f_2)+\theta_1 F_{n_Y}^*(f_1+f_2))\rangle \\
\nonumber =& \; (1-\alpha)^3 \langle \ldots \\
\nonumber & \left[F_{n_X}(f_1)F_{n_Y}(f_2)+\theta_2F_{n_X}(f_1)F_{n_X}(f_2)+\theta_1F_{n_Y}(f_1)F_{n_Y}(f_2)+\theta_1\theta_2F_{n_Y}(f_1)F_{n_X}(f_2)\right](F_{n_X}^*(f_1+f_2)+\theta_1 F_{n_Y}^*(f_1+f_2)) \ldots\\
\nonumber &- \left[F_{n_Y}(f_1)F_{n_X}(f_2)+\theta_1F_{n_Y}(f_1)F_{n_Y}(f_2)+\theta_2F_{n_X}(f_1)F_{n_X}(f_2)+\theta_1\theta_2F_{n_X}(f_1)F_{n_Y}(f_2)\right](F_{n_X}^*(f_1+f_2)+\theta_1 F_{n_Y}^*(f_1+f_2))\ldots\\
\nonumber &\ldots \rangle\\
\nonumber =& \; (1-\alpha)^3\Big\langle(F_{n_X}^*(f_1+f_2)+\theta_1 F_{n_Y}^*(f_1+f_2))\ldots \\
&\cdot \left[F_{n_X}(f_1)F_{n_Y}(f_2)-F_{n_Y}(f_1)F_{n_X}(f_2)+\theta_1\theta_2F_{n_Y}(f_1)F_{n_X}(f_2)-\theta_1\theta_2F_{n_X}(f_1)F_{n_Y}(f_2)\right]\Big\rangle \;.
\label{eq:asymnoise}
\end{align}

\noindent In Eq.~\eqref{eq:asymnoise}, we are able to subtract all noise terms with equal indices, i.e., $(F_{n_X}F_{n_X})$, $(F_{n_Y}F_{n_Y})$. What remains are expressions in $[.]$ with unequal indices only. When expanding this last expression, the final sum must contain only terms with three factors where one index always differs from the other two. By taking the expectation of each term separately, we can factor out the standalone noise term and see that the expectation is equal to zero. Eq.~\eqref{eq:antisym2} thus reduces to: 
\begin{align}
\nonumber B_{[X|YX]}(f_1,f_2)=& \; \alpha^3\beta\langle F_x(f_1)F_y(f_2)F_x^*(f_1+f_2)\rangle \ldots\\
&-\alpha^3\beta\langle F_y(f_1)F_x(f_2)F_x^*(f_1+f_2)\rangle \;.
\end{align}

\noindent Therefore, no noise contributions are present in the antisymmetrized bispectrum.

\subsection{Phase Information Contained in The Antisymmetrized Bispectral Hologram, M1}
\noindent Lastly, we reconceive the phase information contained in the antisymmetrized bispectral hologram. We show that for $M1$ from \cite{Nikias1988}, the composite phase is: 
\begin{align}
\nonumber     \angle(B_{[X|YX]}(f_1,f_2))-\angle(B_{XXX}(f_1,f_2))
 & =\angle(B_{XYX}(f_1,f_2))-\angle(B_{YXX}(f_1,f_2))-\angle(B_{XXX}(f_1,f_2))\\
\nonumber     &= \langle\varphi_{X}(f_1)+\varphi_{Y}(f_2)-\varphi_{X}(f_1+f_2)\rangle - \langle\varphi_{Y}(f_1)+\varphi_{X}(f_2)-\varphi_{X}(f_1 + f_2)\rangle \ldots\\
\nonumber     & \quad - \langle\varphi_{X}(f_1)+\varphi_{X}(f_2)-\varphi_{X}(f_1 + f_2)\rangle\\
    &= \langle \varphi_{Y}(f_2)-2\varphi_{X}(f_2)-\varphi_{Y}(f_1)+\varphi_{X}(f_1+f_2)\rangle \;.
\end{align}
\noindent In the first two phase terms, we see an approximation of the phase difference spectrum across the $f_2$ frequency component. However, a combination of all three frequency indices $f_1,f_2,f_1+f_2$ contribute to TDE outcome. 

\vspace{-2cm}
\end{strip}

\begin{figure*}[htbp]
\centering
\includegraphics[width=.9\linewidth]{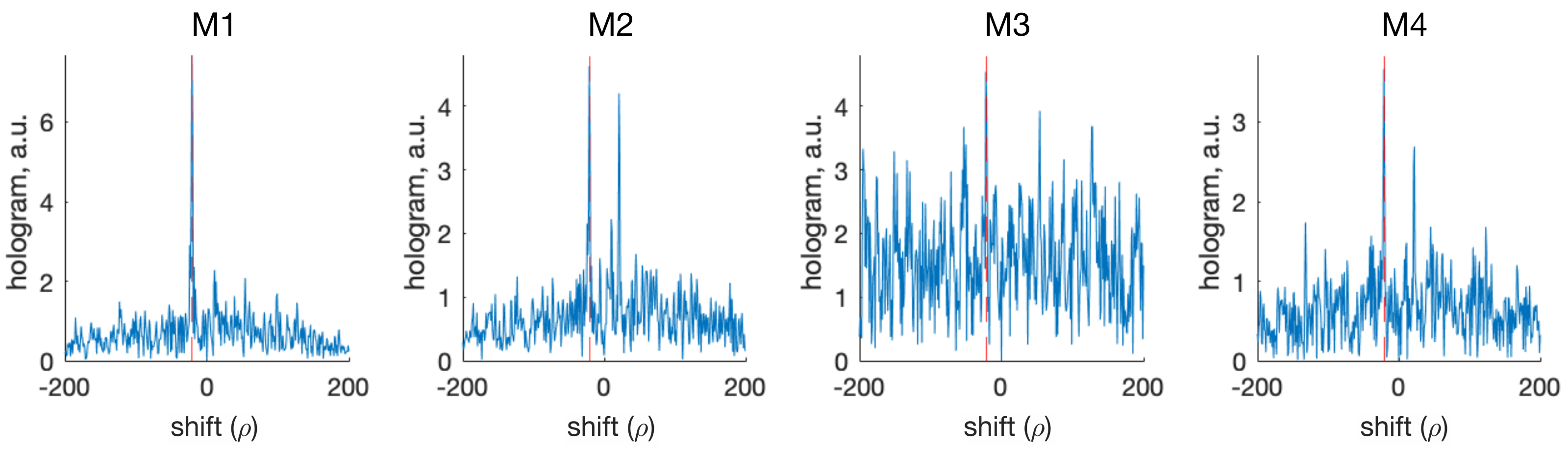}
\caption{Bispectral holograms of antisymmetrized bispectra derived via \textit{Methods I-IV} introduced in \cite{Nikias1988} for exponential signal and mixed exponential noise (SNR $\alpha =.2$). While \textit{M1} produces a hologram with a strong, discernible peak at the underlying delay (marked red), this is not the case for \textit{M2-4}: In the \textit{M3} panel, SNR is poor overall. In the \textit{M2} and \textit{M4} related holograms, an additional peak at the value of the negative delay is visible and can make TDE outcome ambiguous. Thus, to maximize TDE performance in low-SNR environments, we selected \textit{M1} for further analyses.}
\label{supp:m14}
\end{figure*}

\begin{figure*}[htbp]
\centering
\includegraphics[width=.5\linewidth]{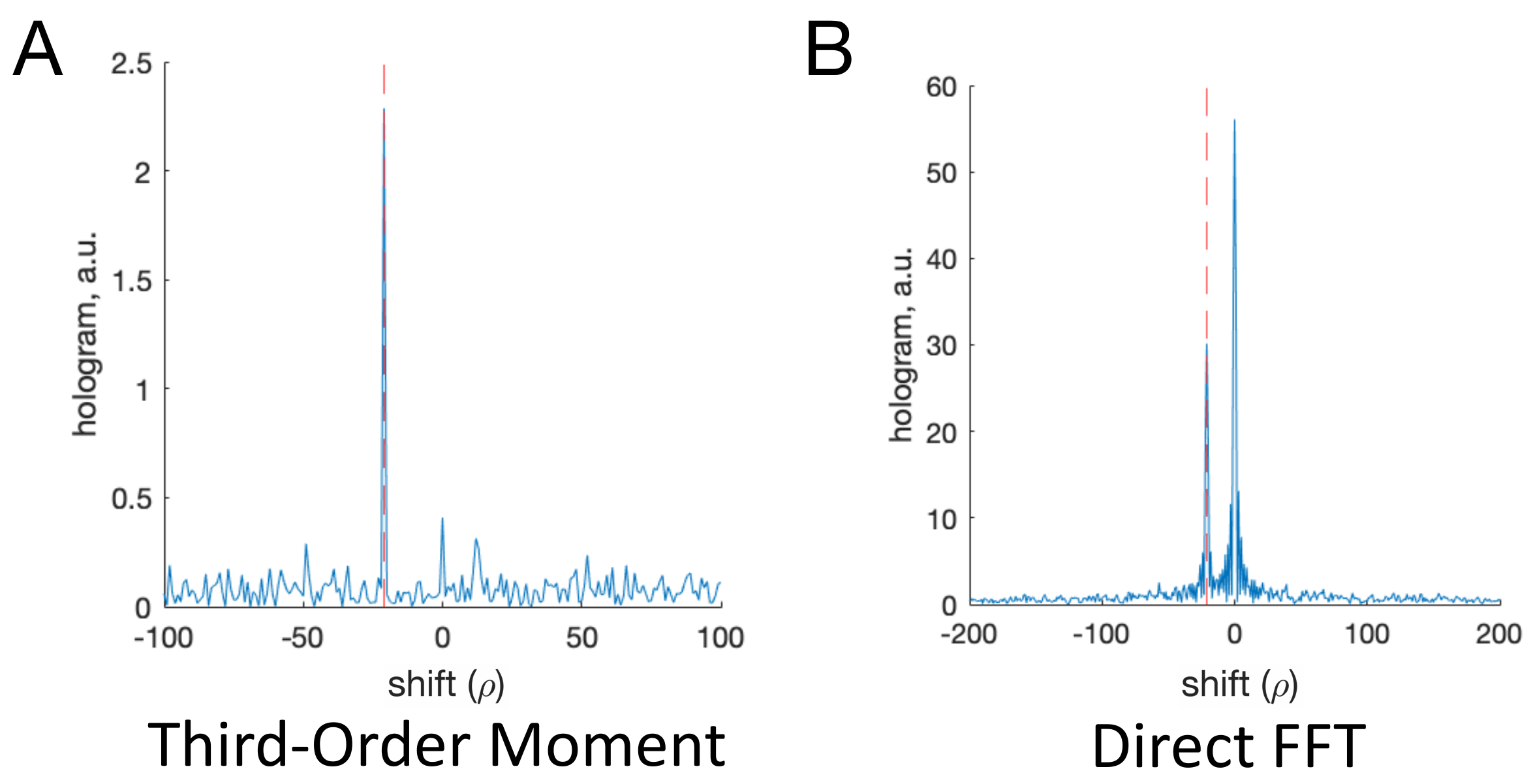}
\caption{Bispectral hologram of two time-delayed time series with exponential signal and mixed Gaussian noise components at SNR $\alpha = .5$. The bispectral hologram was computed A: via the third-order moment or, B: directly via the FFT of the time series. In the former case, no mixed noise influence can be observed, while in the latter case, the interaction of Gaussian noise terms produces an additional peak at $\rho=0$. }
\label{supp:oldvsnewbispec}
\end{figure*}

\begin{figure*}[htbp]
\centering
\includegraphics[width=7in]{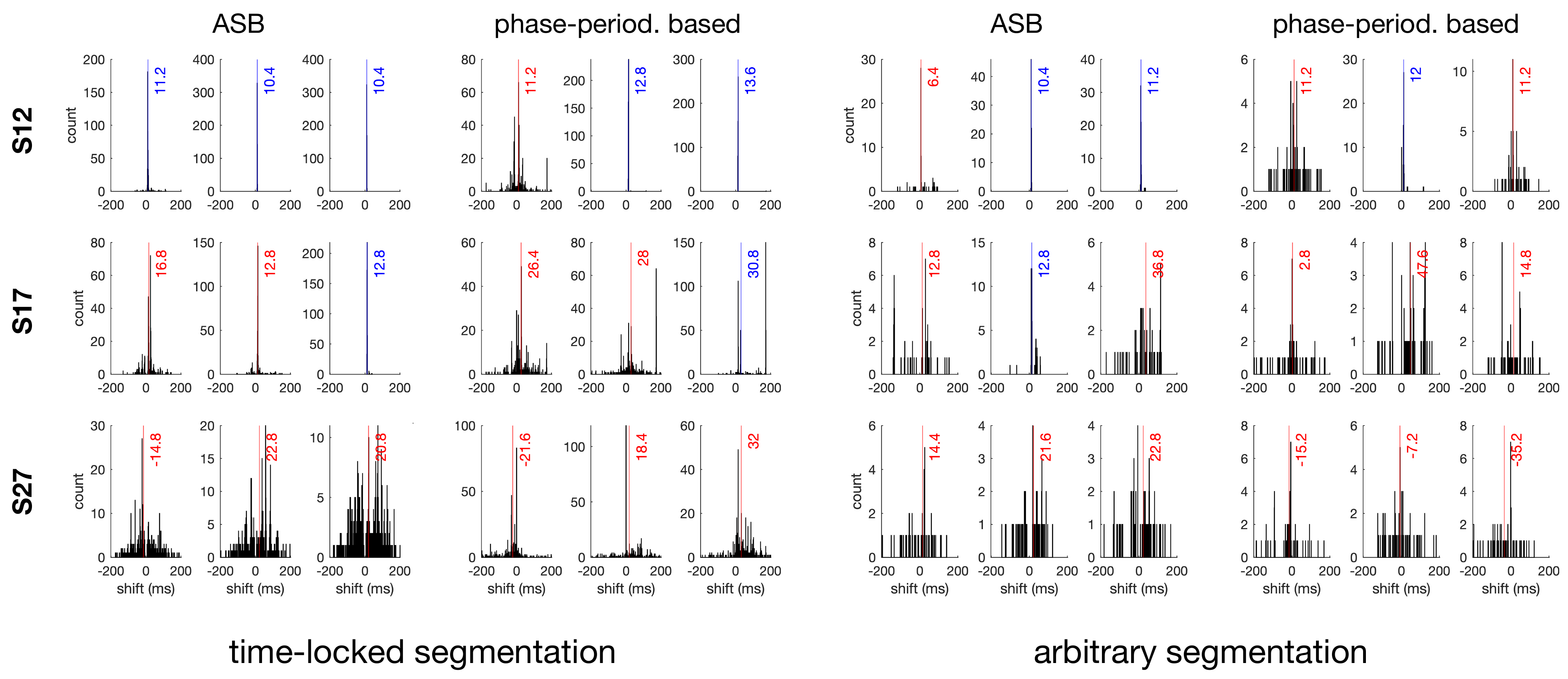}
\caption{Time delay estimation (TDE) between between peripheral and brain electrophysiological recordings. Delays are estimated either on stimulus-locked (left) or arbitrarily segmented (right) epochs in three subjects (S12, S17, and S27). Each plot depicts a total of $n_{\text{boot}}=500$ bootstrapped TD estimates obtained by applying TDE approaches to one of three peripheral-to-EEG (F4, CP4, P4) electrode pairs. Shown are estimates obtained using antisymmetrized bispectral (ASB) and phase-periodicity based TDE approaches. The median of each generated distribution is marked and color coded: Blue indicates acceptance of the TDE estimate according to our filtering approach with a chosen confidence interval width of 95\%, red indicates rejection of the trial. Accepted time delay estimates for stimulus-locked and arbitrary segmentation analyses are closely matching.}

\label{fig:realdatatde}
\end{figure*}

\end{document}